\documentclass[aps,prresearch,reprint,superscriptaddress,longbibliography,floatfix]{revtex4-2}
\usepackage{graphicx}
\usepackage{amsmath,amssymb,amsfonts,bm}
\usepackage{epsfig}
\usepackage{epstopdf}
\usepackage{dcolumn}
\usepackage{grffile}
\usepackage{verbatim}
\usepackage{mathrsfs}
\usepackage{appendix}
\usepackage{xcolor}
\usepackage{tikz}
\usepackage{txfonts}
\usepackage[normalem]{ulem}
\usepackage[colorlinks=true,linkcolor=blue,citecolor=blue, urlcolor=blue]{hyperref}
\usetikzlibrary{arrows.meta,calc}

\definecolor{alizarin}{rgb}{0.82, 0.1, 0.26}

\begin{document}
	
	\title{Intrinsic Floquet Generation and $1/I$ Quantum Oscillations in a Sliding Charge-Density Wave}
	\date{\today}
	
	\author{Yi Zhou}
	\email{yizhou@iphy.ac.cn}
	\affiliation{Institute of Physics, Chinese Academy of Sciences, Beijing 100190, China}
	
\begin{abstract}
	Recent experiments [Phys. Rev. B \textbf{109}, 245123 (2024)] revealed striking inverse-current ($1/I$) quantum oscillations in quasi-one-dimensional charge-density-wave (CDW) insulators and proposed an intrinsic Floquet sideband mechanism arising from the sliding condensate. Here we develop the complete theoretical framework underlying this proposal. We provide an exact Floquet diagonalization of the uniformly sliding CDW problem, yielding split gap edges and a ladder of Floquet sidebands with explicit unitary transformation and spectral functions. Using this exact solution, we formulate weak-probe tunneling spectroscopy and demonstrate that the local Floquet spectrum naturally yields $1/I$ oscillations as successive sideband edges cross a fixed contact chemical potential. Matching the observed oscillation period to theory reveals that the macroscopic current must percolate through a highly localized coherent filament, with effective channel number $N_{\mathrm{eff}} \sim 480$, nearly two orders of magnitude smaller than the geometric chain count $N_{\mathrm{geom}} \sim 3 \times 10^4$. This filamentary confinement is essential: achieving the required sliding frequency uniformly across the bulk would demand prohibitively large currents and induce thermal dephasing. Furthermore, using a segmented multiterminal model, we show that inelastic phase-slip dephasing near the contacts explains the experimentally observed suppression of oscillation visibility on outer voltage probes. Finally, we contrast the persistent-current-driven multiterminal geometry with a homogeneous voltage-biased two-terminal reference calculation. Our results establish a rigorous nonequilibrium transport framework for the observed $1/I$ oscillations and highlight a universal spatial-to-temporal conversion mechanism in which the insulating gap protects Floquet coherence, offering a design principle for intrinsically driven quantum devices.
\end{abstract}
	
	\maketitle
	
\section{Introduction}
\label{sec:intro}

Macroscopic quantum states such as superconductivity and charge density waves (CDWs) offer unique platforms for exploring emergent quantum phenomena and developing next-generation quantum materials~\cite{Frohlich54,Peierls,CDW88,Monceau2012}. A CDW is a spontaneously formed, spatially periodic modulation of the electronic charge density and lattice coordinates that naturally arises in quasi-one-dimensional (quasi-1D) conductors~\cite{Frohlich54,Peierls}, with related considerations applying to spin density waves~\cite{SDW60,SDW62,SDW94}. Above a critical depinning threshold, the CDW condensate can slide collectively, leading to highly nonlinear charge transport~\cite{LeeRiceAnderson1974,Monceau1976,FukuyamaLee1978,LeeRice1979,FlemingGrimes1979,Bardeen1979,Monceau2012}.

A particularly striking manifestation of macroscopic CDW coherence was recently reported by Le \emph{et al.}~\cite{Le24}, who observed robust quantum oscillations periodic in inverse applied current ($1/I$) in quasi-1D CDW insulators (TaSe$_4$)$_2$I and TaS$_3$. These oscillations appear in the Fr\"ohlich conductivity regime where the depinned condensate slides coherently, and disappear once the oscillation index reaches unity, mirroring the quantum limit of Shubnikov--de Haas oscillations. Crucially, the same work proposed that this phenomenon originates from sliding-driven inherent Floquet sidebands: once the CDW condensate slides uniformly, its phase $\phi(t) = \phi_0 - \Omega t$ becomes explicitly time dependent, so the condensate itself generates a periodic drive. This intrinsic spatial-to-temporal conversion represents a fundamental departure from the more familiar Floquet settings, which typically require external microwave fields or ultrafast laser pulses~\cite{Grifoni,Platero04,Kohler05,Bukov2015,OkaKitamura2019,RudnerLindner2020}.

Reference~\cite{Le24} sketched the basic Floquet mechanism and derived the leading periodicity relation $\Delta(1/I) = h/(2eN_{\mathrm{eff}}\epsilon)$, but explicitly called for a well-developed theoretical model to address the full phenomenology. Several essential questions were left open: (i) Is the sliding-CDW Floquet problem exactly solvable, and what is the structure of its complete spectrum and Green's functions? (ii) How does the local Floquet sideband ladder manifest in different measurement geometries---weak tunneling spectroscopy versus contacted transport? (iii) Why does the macroscopic terminal current produce sideband spacings far larger than naive geometric estimates would suggest? (iv) Why is the $1/I$ oscillation strongly visible on inner voltage probes but heavily suppressed on outer terminal pairs? The present work develops the complete theoretical framework needed to address these questions.

Our principal contributions are as follows. First, we provide an exact Floquet diagonalization of the uniformly sliding CDW, with explicit unitary transformation, quasienergy spectrum, eigenstates, and time-averaged spectral functions (Sec.~\ref{sec:floquet}). Second, we formulate a complete Schwinger-Keldysh nonequilibrium framework that distinguishes weak-probe tunneling spectroscopy from contacted two-terminal transport, and we identify the local time-averaged spectral function as the singular structure governing observed oscillations (Sec.~\ref{sec:spectroscopy}). Third, by quantitatively matching the observed oscillation period to theory, we deduce that the macroscopic current must percolate through a highly localized coherent filament, with effective channel number $N_{\mathrm{eff}} \sim 480$ nearly two orders of magnitude smaller than the geometric chain count $N_{\mathrm{geom}} \sim 3 \times 10^4$. This filamentary confinement is not incidental but essential: it allows the system to reach observable sliding frequencies without prohibitive Joule heating that would destroy Floquet coherence. Fourth, we introduce a segmented multiterminal transport model that explains the experimentally observed visibility hierarchy: phase-slip dephasing near the current-injecting contacts confines coherent oscillations to the inner segments, accounting for why $V_{2-3}^{1-4}$ shows clean $1/I$ oscillations while the raw outer-terminal voltage $V_{1-4}^{1-4}$ requires background subtraction (Sec.~\ref{sec:transport}). Finally, we contrast this experimentally relevant geometry with a homogeneous voltage-biased two-terminal reference calculation to clarify how the same Floquet sideband ladder projects onto different observables.

Beyond the specific CDW context, the framework underscores a critical material prerequisite for observing such macroscopic Floquet coherence. In two-dimensional CDWs with residual Fermi surfaces, low-energy electron-electron scattering produces rapid dephasing that smears out delicate sideband structure. By contrast, the complete excitation gap of a quasi-1D CDW insulator exponentially suppresses these scattering channels, protecting the coherent Floquet dynamics. This gap-protected coherence offers a design principle for future quantum devices, allowing highly tunable high-frequency drives to be generated locally via simple dc currents without the need for external microwave architecture.

\begin{figure}[tb]
	\centering
	\includegraphics[width=0.92\linewidth]{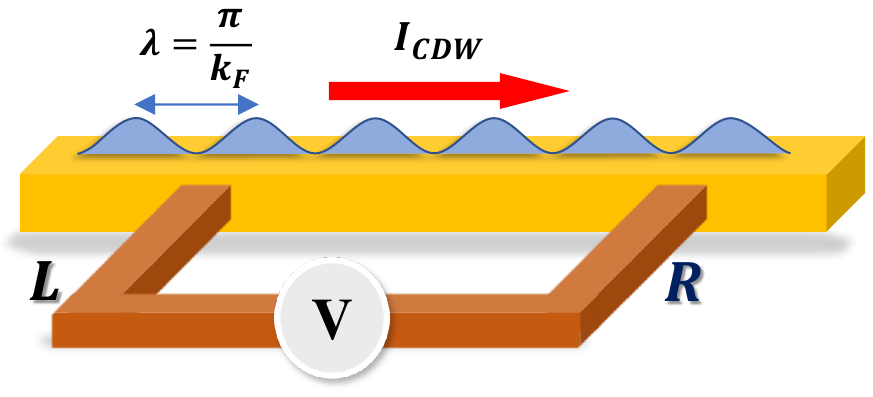}
	\caption{\label{Fig:device} Two-terminal sliding-CDW reference geometry: a finite sliding CDW region coupled to left and right metallic leads. This homogeneous voltage-biased geometry serves as a theoretical reference limit (Sec.~\ref{sec:transport}) and is distinct both from the weak single-probe spectroscopy of Sec.~\ref{sec:spectroscopy} and from the persistent-current-driven multiterminal experiment of Ref.~\cite{Le24} sketched in Fig.~\ref{Fig:segmentedSchematic}.}
\end{figure}

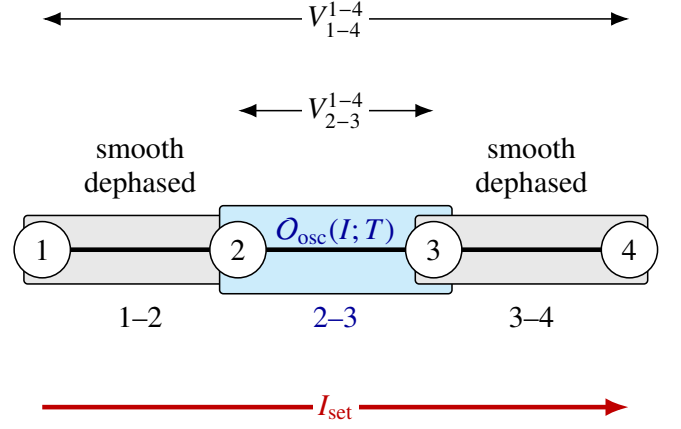
\begin{figure}[tb]
	\centering
	\resizebox{\linewidth}{!}{%
		\begin{tikzpicture}[
			x=1cm,y=1cm,scale=1.12,transform shape,
			terminal/.style={circle, draw=black, fill=white, minimum size=4.6mm, inner sep=0pt, font=\scriptsize},
			outerseg/.style={draw=black, fill=gray!18, rounded corners=1pt},
			innerseg/.style={draw=black, fill=cyan!18, rounded corners=1pt},
			meas/.style={thin, <->, >=Latex},
			curr/.style={very thick, -{Latex[length=2.2mm]}},
			lab/.style={font=\scriptsize, align=center}
			]
			\coordinate (a) at (0,0);
			\coordinate (b) at (1.60,0);
			\coordinate (c) at (3.20,0);
			\coordinate (d) at (4.80,0);
			\path[outerseg] (-0.15,-0.28) rectangle (1.75,0.28);
			\path[innerseg] (1.45,-0.36) rectangle (3.35,0.36);
			\path[outerseg] (3.05,-0.28) rectangle (4.95,0.28);
			\draw[very thick] (a) -- (d);
			\node[terminal] (t1) at (a) {1};
			\node[terminal] (t2) at (b) {2};
			\node[terminal] (t3) at (c) {3};
			\node[terminal] (t4) at (d) {4};
			\node[lab] at (0.80,-0.55) {1--2};
			\node[lab,text=blue!60!black] at (2.40,-0.55) {2--3};
			\node[lab] at (4.00,-0.55) {3--4};
			\node[lab] at (0.80,0.67) {smooth\\dephased};
			\node[lab,text=blue!60!black] at (2.40,0.15) {$\mathcal{O}_{\mathrm{osc}}(I;T)$};
			\node[lab] at (4.00,0.67) {smooth\\dephased};
			\draw[meas] ($(t2.north)+(0,0.90)$) -- node[lab,fill=white,inner sep=1pt] {$V_{2-3}^{1-4}$} ($(t3.north)+(0,0.90)$);
			\draw[meas] ($(t1.north)+(0,1.65)$) -- node[lab,fill=white,inner sep=1pt] {$V_{1-4}^{1-4}$} ($(t4.north)+(0,1.65)$);
			\draw[curr,red!75!black] ($(t1.south)+(0,-1.05)$) -- node[lab,fill=white,inner sep=1pt,text=red!75!black] {$I_{\mathrm{set}}$} ($(t4.south)+(0,-1.05)$);
		\end{tikzpicture}
	}
	\caption{\label{Fig:segmentedSchematic} Experiment-motivated segmented model for a persistent-current-driven multiterminal device~\cite{Le24}. One imposed current $I_{\mathrm{set}}$ flows through three serial segments. The central segment $2$-$3$ hosts the coherent Floquet kernel $\mathcal{O}_{\mathrm{osc}}(I;T)$ [defined in Eq.~\eqref{eq:segmentedKernel} of Appendix~\ref{sec:segmented-transport}], whereas the outer segments contribute smooth, effectively dephased voltage drops due to phase-slip-induced inelastic scattering near the contacts. The experimentally relevant inner- and outer-terminal voltages are indicated above the device.}
\end{figure}
	
\section{Sliding CDW and exact Floquet solution}
\label{sec:floquet}

\subsection{Static and sliding order parameter}

At low energies the CDW amplitude mode is gapped, so we retain only the phase degree of freedom and write
\begin{equation}
	\Delta(x,t)=\Delta e^{i\phi(x,t)},
\end{equation}
with $\Delta>0$. Once a dc current exceeds the depinning threshold, the phase evolves linearly in time, $\phi(t)=\phi_{0}-\Omega t$, so that the density modulation becomes
\begin{equation}
	\rho(x,t)=\rho_{0}+\rho_{1}\cos(Qx-\Omega t+\phi_{0}),
\end{equation}
where $Q=2k_{F}$. Following Ref.~\cite{Le24}, the sliding frequency is fixed by the condensate current per chain, $\Omega=\pi j_{\mathrm{CDW}}/e$, or equivalently $j_{\mathrm{CDW}}=-(e/\pi)\dot{\phi}$~\cite{CDW88}. In a realistic macroscopic device, however, the total imposed current $I$ need not flow uniformly through the entire geometric cross-section. If the sliding state depins along a preferred path due to inhomogeneous pinning, the current per chain in this active region is $j_{\mathrm{CDW}} = I / N_{\mathrm{eff}}$, where $N_{\mathrm{eff}}$ denotes the effective number of chains participating in the percolating coherent filament (a quantitative analysis of this filament is given in Sec.~\ref{sec:spectroscopy}). The intrinsic drive frequency is thus directly tied to the macroscopic terminal current via
\begin{equation}\label{eq:Omega-I}
	\Omega = \frac{\pi I}{e N_{\mathrm{eff}}}.
\end{equation}
The order parameter is therefore periodic in time even without external irradiation. Throughout the paper, $\Omega$ denotes the angular sliding frequency and $\hbar\Omega$ is the corresponding photon-like energy quantum of the periodic drive.

\subsection{Two-branch model and exact Floquet spectrum}

To model a single chain within this coherent filament, we retain a single band linearized around the two Fermi points, labeling the right-moving ($k>0$) and left-moving ($k<0$) branches by $r=+1$ and $r=-1$, respectively. This linearized two-branch approximation is valid as long as $v_{F}k_{F}\gg\Delta$. The mean-field Hamiltonian reads
\begin{equation}\label{eq:HCDW}
	H_{\mathrm{CDW}}=\sum_{\mathbf{k},\sigma}\varepsilon_{\mathbf{k}}f_{\mathbf{k}\sigma}^{\dagger}f_{\mathbf{k}\sigma}+\frac{1}{2}\sum_{\mathbf{k},\sigma}\left(\Delta e^{i r\phi}f_{\mathbf{k}+r\mathbf{Q},\sigma}^{\dagger}f_{\mathbf{k},\sigma}+\mathrm{H.c.}\right),
\end{equation}
where $f_{\mathbf{k}\sigma}^{\dagger}$ creates and $f_{\mathbf{k}\sigma}$ annihilates an electron with momentum $\mathbf{k}$ and spin $\sigma$, and $\phi=\phi_{0}-\Omega t$. We further assume $\hbar\Omega\ll\Delta$, so that the sliding motion remains within the same low-energy two-branch description. If $\mathbf{k}$ lies on the $r=\pm1$ branch, then $\mathbf{k}+r\mathbf{Q}$ lies on the opposite branch. Introducing the spinor field $\psi_{\mathbf{k}\sigma}=\left(f_{\mathbf{k},\sigma},f_{\mathbf{k}+r\mathbf{Q},\sigma}\right)^{T}$, the Hamiltonian takes the compact form
\begin{equation}
	H_{\mathrm{CDW}} = \frac{1}{2}\sum_{\mathbf{k},\sigma}\psi^{\dagger}_{\mathbf{k}\sigma}H_{\mathbf{k}}\psi_{\mathbf{k}\sigma},
\end{equation}
with the $2\times 2$ matrix
\begin{equation}
	H_{\mathbf{k}}= \left[\begin{array}{cc}\varepsilon_{\mathbf{k}} & \Delta e^{-ir\phi}\\\Delta e^{ir\phi} & \varepsilon_{\mathbf{k}+r\mathbf{Q}}\end{array}\right].
\end{equation}

In the static limit $\phi(t)=\phi_{0}$, one recovers the usual gapped two-branch dispersion. For a sliding CDW, however, the explicit time dependence of $H_{\mathbf{k}}$ prevents a direct static diagonalization. We therefore solve the time-dependent Schr\"{o}dinger equation using the single-particle Floquet Hamiltonian~\cite{Floquet,Shirley,Sambe}
\begin{equation}
	H_{\mathbf{k}}^{F}=H_{\mathbf{k}}-i\hbar\frac{\partial}{\partial t}.
\end{equation}

The Floquet Hamiltonian $H_{\mathbf{k}}^{F}$ admits exact diagonalization by the time-dependent unitary matrix
\begin{equation}\label{eq:Sk}
	S_{\mathbf{k}}(t) = \left[\begin{array}{cc} e^{-in\Omega t} & 0 \\ 0 & e^{-i(n+r)\Omega t} \end{array}\right]\left[\begin{array}{cc} u_{\mathbf{k}+} & u_{\mathbf{k}-} \\ v_{\mathbf{k}+} & v_{\mathbf{k}-} \end{array}\right],
\end{equation}
where the eigenvectors of the static effective problem are
\begin{equation}\label{eq:uvk}
	\left(\begin{array}{c}u_{\mathbf{k}}\\v_{\mathbf{k}}\end{array}\right)_{+} = \left[\begin{array}{c}\cos\frac{\theta_{\mathbf{k}}}{2}\\\sin\frac{\theta_{\mathbf{k}}}{2}e^{ir\phi_{0}}\end{array}\right],\quad \left(\begin{array}{c}u_{\mathbf{k}}\\v_{\mathbf{k}}\end{array}\right)_{-} = \left[\begin{array}{c}-\sin\frac{\theta_{\mathbf{k}}}{2}\\\cos\frac{\theta_{\mathbf{k}}}{2}e^{ir\phi_{0}}\end{array}\right].
\end{equation}
The two labels ``$+$'' and ``$-$'' refer to the sign of $\mathrm{sgn}(\varepsilon_{\mathbf{k}}-\varepsilon_{F})=\pm 1$; the partner state at $\mathbf{k}+r\mathbf{Q}$ carries the opposite sign. Diagonalization yields
\begin{equation}
	S_{\mathbf{k}}(t)^{\dagger}H_{\mathbf{k}}^{F}S_{\mathbf{k}}(t)=\mathcal{E}_{\mathbf{k}}\sigma_3+\left[a_{\mathbf{k}}+\left(n+\frac{r}{2}\right)\hbar\Omega\right]\sigma_0.
\end{equation}
A detailed derivation of this exact two-level Floquet diagonalization is given in Appendix~\ref{sec:2L}. Here we have introduced
\begin{subequations}
	\begin{equation}
		a_{\mathbf{k}}=a_{\mathbf{k}+r\mathbf{Q}}=\frac{\varepsilon_{\mathbf{k}}+\varepsilon_{\mathbf{k}+r\mathbf{Q}}}{2},
	\end{equation}
	\begin{equation}
		\mathcal{E}_{\mathbf{k}} = \mathcal{E}_{\mathbf{k}+r\mathbf{Q}} = \sqrt{\left(\frac{\varepsilon_{\mathbf{k}}-\varepsilon_{\mathbf{k}+r\mathbf{Q}}-r\hbar\Omega}{2}\right)^2+\Delta^2},
	\end{equation}
	\begin{equation}
		\cos\theta_{\mathbf{k}} =-\cos\theta_{\mathbf{k}+r\mathbf{Q}}= \frac{\varepsilon_{\mathbf{k}}-\varepsilon_{\mathbf{k}+r\mathbf{Q}}-r\hbar\Omega}{2\mathcal{E}_{\mathbf{k}}}.
	\end{equation}
\end{subequations}
The quasienergies---the eigenvalues of $H_{\mathbf{k}}^{F}$---are
\begin{equation}\label{eq:Ek}
	E_{\mathbf{k}n} = \frac{\varepsilon_{\mathbf{k}}+\varepsilon_{\mathbf{k}+r\mathbf{Q}}}{2} + \left(n+\frac{r}{2}\right)\hbar\Omega + \text{sgn}\left(\varepsilon_{\mathbf{k}}-\varepsilon_{F}\right)\mathcal{E}_{\mathbf{k}},
\end{equation}
where $n$ is an integer indexing the Floquet sideband. The corresponding eigenstates are
\begin{equation}\label{eq:kn}
	|\mathbf{k},n\rangle=e^{-in\Omega t}\left(u_{\mathbf{k}}|\mathbf{k}\rangle+v_{\mathbf{k}}e^{-ir\Omega t}|\mathbf{k}+r\mathbf{Q}\rangle\right).
\end{equation}
Equivalently, defining the time-dependent creation operator
\begin{equation}\label{eq:dk}
	d_{\mathbf{k}\sigma}^{\dagger} = u_{\mathbf{k}}f_{\mathbf{k}\sigma}^{\dagger}+v_{\mathbf{k}}e^{-ir\Omega t}f_{\mathbf{k}+r\mathbf{Q},\sigma}^{\dagger},
\end{equation}
the operators $d_{\mathbf{k}\sigma}^{\dagger}$ create Floquet quasiparticles with quasienergies $E_{\mathbf{k}0}$. This exact solution provides the full Floquet spectrum that underlies the qualitative sideband picture sketched in Ref.~\cite{Le24}.

\subsection{Green's functions and Floquet spectral functions}

For a Hamiltonian periodic in time with period $2\pi/\Omega$, the double-time Green's function satisfies $F(t_1,t_2)=F(t_1+2\pi/\Omega,t_2+2\pi/\Omega)$. In Wigner coordinates $\tau=(t_1+t_2)/2$ and $t=t_1-t_2$, its Floquet representation is
\begin{equation}\label{eq:FloquetG}
	F(t_1,t_2)=\sum_{n}\int\frac{d\omega}{2\pi}e^{-i\omega t}e^{in\Omega \tau}F(\omega,n),
\end{equation}
where $t_{1}=\tau+t/2$ and $t_{2}=\tau-t/2$.

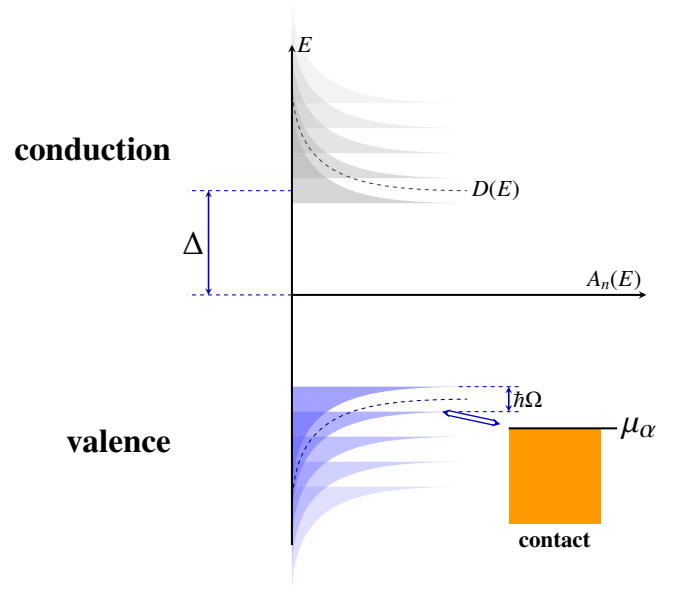
\begin{figure}[tb]
	\centering
	\resizebox{\linewidth}{!}{%
		\begin{tikzpicture}[
			x=1.1cm,y=1.1cm,
			>=stealth
			]
			\foreach \i in {4, 3, 2, 1, 0} {
				\pgfmathsetmacro\op{0.55 - \i*0.1}
				\pgfmathsetmacro\ypos{2.2 + \i*0.6}
				\fill[gray!60, opacity=\op] 
				(0, \ypos+2.5) .. controls (0.1, \ypos+0.3) and (1.5, \ypos+0.02) .. (4.2, \ypos) -- (0, \ypos) -- cycle;
			}
			\foreach \i in {4, 3, 2, 1, 0} {
				\pgfmathsetmacro\op{0.6 - \i*0.1}
				\pgfmathsetmacro\ypos{-2.2 - \i*0.6}
				\fill[blue!60, opacity=\op] 
				(0, \ypos-2.5) .. controls (0.1, \ypos-0.3) and (1.5, \ypos-0.02) .. (4.2, \ypos) -- (0, \ypos) -- cycle;
			}
			\draw[thick, dashed, darkgray] 
			(0, 5.0) .. controls (0.1, 2.8) and (1.5, 2.52) .. (4.2, 2.5) 
			node[right, font=\LARGE\bfseries, text=black] {$D(E)$};
			\draw[thick, dashed, blue!50!black] 
			(0, -5.0) .. controls (0.1, -2.8) and (1.5, -2.52) .. (4.2, -2.5);
			\draw[->, line width=1.5pt] (0, -6) -- (0, 6) node[right, font=\LARGE\bfseries] {$E$};
			\draw[->, line width=1.5pt] (0, 0) -- (8.5, 0) node[above left, font=\LARGE\bfseries] {$A_n(E)$};
			\draw[dashed, thick, blue!70!black] (0, 2.5) -- (-2.5, 2.5);
			\draw[dashed, thick, blue!70!black] (0, 0) -- (-2.5, 0);
			\draw[<->, line width=1.2pt, blue!70!black] (-2.0, 0) -- (-2.0, 2.5) 
			node[midway, left, font=\Huge\bfseries, text=black] {$\Delta$};
			\node[left, font=\Huge\bfseries] at (-2.8, 3.5) {conduction};
			\node[left, font=\Huge\bfseries] at (-2.8, -3.5) {valence};
			\draw[dashed, thick, blue!70!black] (4.0, -2.2) -- (5.5, -2.2);
			\draw[dashed, thick, blue!70!black] (4.0, -2.8) -- (5.5, -2.8);
			\draw[<->, line width=1.2pt, blue!70!black] (5.2, -2.2) -- (5.2, -2.8) 
			node[midway, right, font=\LARGE\bfseries, text=black] {$\hbar\Omega$};
			\fill[orange!80!yellow] (5.2, -5.5) rectangle (7.4, -3.2);
			\draw[line width=1.5pt, black] (5.2, -3.2) -- (7.8, -3.2) 
			node[right, font=\Huge\bfseries] {$\mu_\alpha$};
			\node[below, font=\LARGE\bfseries] at (6.3, -5.6) {contact};
			\draw[<->, double, double distance=2.5pt, line width=1pt, blue!70!black] 
			(3.6, -2.8) -- (5.0, -3.1);
		\end{tikzpicture}
	}
	\caption{\label{Fig:DOS} Integral Floquet spectral function $A_{n}(E)$ of the sliding CDW. The dashed curve shows the density of states $D(E)$ of the corresponding static CDW state. Once the CDW slides, the square-root gap edges split by $\pm\hbar\Omega/2$ and form a ladder of sidebands separated by $\hbar\Omega$. The double arrow indicates tunneling between the CDW and a metallic contact at chemical potential $\mu_{\alpha}$.}
\end{figure}

For the sliding CDW, the retarded and advanced Green's functions~\cite{Mahan} in the Floquet basis are
\begin{equation}\label{eq:G}
	g^{R/A}_{\mathbf{k}}(\omega,n) =\frac{1}{\omega\pm i0^{+}-E_{\mathbf{k}n}},
\end{equation}
where Eq.~\eqref{eq:kn} defines the Floquet eigenbasis. The integral Floquet spectral function~\cite{Uhrig} is
\begin{subequations}
	\begin{equation}
		A_{n}(E):=-\frac{1}{\pi}\mathrm{Im}\sum_{\mathbf{k}}g_{\mathbf{k}}^{R}(E,n)=A_{0}(E-n\hbar\Omega),
	\end{equation}
	where $A_{0}(E)$ is the time-averaged density of states (DOS), derived in Appendix~\ref{sec:An}:
	\begin{equation}
		\begin{split}
			A_{0}(E)&=\frac{1}{2}\left[D\left(E+\frac{\hbar\Omega}{2}\right)+D\left(E-\frac{\hbar\Omega}{2}\right)\right],\\
			D(E)&=\frac{N(0)\left|E-\varepsilon_{F}\right|}{\sqrt{\left(E-\varepsilon_{F}\right)^2-\Delta^2}}\theta\left(\left|E-\varepsilon_{F}\right|-\Delta\right).
		\end{split}
	\end{equation}
\end{subequations}
Here $D(E)$ is the DOS of the static CDW state with $\Omega=0$, and $N(0)=(\pi\hbar v_{F})^{-1}$ is the Fermi-level DOS of the noninteracting band. Figure~\ref{Fig:DOS} illustrates how the static square-root edges split by $\pm\hbar\Omega/2$ once the CDW slides, generating the Floquet sideband structure central to the transport phenomenology developed below.

The exact Floquet solution just obtained fixes the spectrum of the isolated sliding CDW. Once metallic contacts are attached, however, one must distinguish this diagonal Floquet basis from the physical electron basis created by $\{f_{\mathbf{k}\sigma}^{\dagger}\}$, in which the tunneling Hamiltonian is time independent. We therefore use the exact solution as the building block for the physical-basis Green's functions: $\mathbf{\tilde{g}}$ for the isolated sliding CDW and $\mathbf{\tilde{G}}$ for the contacted system. The explicit basis transformation is summarized in Appendix~\ref{sec:noninteracting-basis}.
	
\section{Single-contact tunneling spectroscopy}
\label{sec:spectroscopy}

\begin{figure}[tb]
	\centering
	\resizebox{\linewidth}{!}{%
		\begin{tikzpicture}[
			x=1cm,y=1cm,
			cdw/.style={draw=black, fill=cyan!14, rounded corners=2pt},
			env/.style={draw=black, fill=gray!14, rounded corners=2pt},
			lab/.style={font=\scriptsize, align=center},
			drive/.style={very thick, -{Latex[length=2.1mm]}, red!75!black},
			tunnel/.style={very thick, <->, >=Latex, blue!65!black},
			guide/.style={thin, dashed, gray!70}
			]
			\path[env] (-0.85,0.08) rectangle (0.02,0.48);
			\path[cdw] (0,0) rectangle (6.10,0.56);
			\path[env] (6.08,0.08) rectangle (6.95,0.48);
			\draw[thick, blue!60!black, smooth, samples=120, domain=0.18:5.92] plot (\x,{0.28+0.11*sin(720*\x)});
			\draw[guide] (2.55,-0.04) rectangle (3.65,0.60);
			\node[lab] at (1.50,0.68) {Sliding CDW};
			\node[lab,text=blue!60!black] at (3.10,-0.34) {Sampled local\\Floquet spectrum};
			\draw[drive] (0.40,-0.74) -- node[lab,fill=white,inner sep=1pt,text=red!75!black] {Collective sliding, $\Omega(I)$} (5.70,-0.74);
			\node[lab] at (-0.14,0.88) {External drive\\and relaxation};
			\node[lab] at (6.22,0.88) {External drive\\and relaxation};
			\draw[black, fill=orange!18] (2.52,2.18) rectangle (3.68,2.58);
			\draw[black, fill=orange!18] (2.76,2.18) -- (3.44,2.18) -- (3.10,1.10) -- cycle;
			\node[lab] at (3.10,2.38) {Probe $P$};
			\draw[tunnel] (3.10,0.98) -- node[right=1pt,lab] {Weak tunneling\\$H_{T,P}$} (3.10,0.68);
			\draw[thin, <->, >=Latex] (3.96,2.38) -- node[lab,fill=white,inner sep=1pt] {$V_{P}$} (3.96,0.28);
			\node[lab] at (4.52,2.38) {$\mu_{P}$};
			\draw[thin, -{Latex[length=1.8mm]}] (4.08,1.54) -- (5.10,1.54);
			\node[lab] at (5.74,1.54) {$\langle J_{P}\rangle$};
		\end{tikzpicture}
	}
	\caption{\label{Fig:singleProbeSetup} Weak single-contact tunneling geometry. A metallic probe $P$ is weakly tunnel-coupled to a local region of the sliding CDW and biased by $V_{P}$ relative to the local CDW potential. The probe current $\langle J_{P}\rangle$ samples the local Floquet spectrum, while the sliding steady state and frequency $\Omega(I)$ are maintained elsewhere by the external dc drive and intrinsic relaxation of the CDW.}
\end{figure}
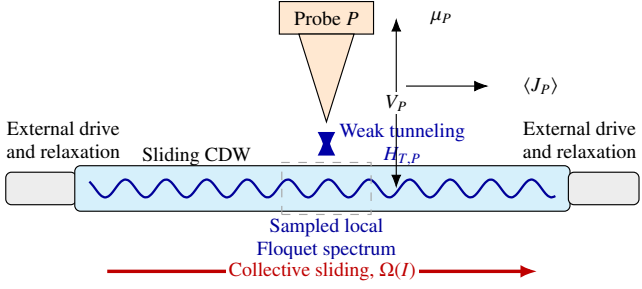

We now develop the simplest measurement geometry in which the Floquet sideband ladder of Sec.~\ref{sec:floquet} can be directly observed: weak single-contact tunneling spectroscopy. A metallic probe $P$ is tunnel-coupled to the sliding CDW, while the sliding steady state itself is maintained by the external dc drive and intrinsic relaxation of the CDW elsewhere in the device. This is spectroscopy rather than two-terminal transport: the probe samples the local Floquet spectrum but does not set the conserved through-current. The geometry is sketched in Fig.~\ref{Fig:singleProbeSetup}. While Ref.~\cite{Le24} briefly discussed sideband-assisted tunneling at the contact, we provide here the complete nonequilibrium formulation and use it to deduce the physical nature of the sliding state through quantitative matching to experiment.

\subsection{Setup and time-averaged probe current}
The full Hamiltonian for this setup is
\begin{subequations}
	\begin{equation}
		H=H_{\mathrm{CDW}}+H_{P}+H_{T,P},
	\end{equation}
	\begin{equation}
		H_{P}=\sum_{k}\epsilon_{kP}c_{kP}^{\dagger}c_{kP},\quad H_{T,P}=\sum_{k,p,\sigma}V_{kP,p}c_{kP}^{\dagger}f_{p\sigma}+\mathrm{H.c.},
	\end{equation}
\end{subequations}
where $p$ labels physical single-particle states in the CDW region. Because $H_{T,P}$ is time independent in the $f$ basis, the probe self-energy is stationary and depends only on the relative time $t_1-t_2$ (see Appendix~\ref{sec:lead-self-energy}). The time-averaged probe current then reads
\begin{equation}\label{eq:JPavg}
	\langle J_{P}\rangle = -\frac{e}{\hbar}\mathrm{Im}\int\frac{d\omega}{2\pi}\mathrm{tr}\left\{\mathbf{\Gamma}^{P}(\omega)\left[\mathbf{\tilde{G}}^{<}(\omega,0)+2f_{P}(\omega)\mathbf{\tilde{G}}^{R}(\omega,0)\right]\right\},
\end{equation}
where $\mathbf{\Gamma}^{P}(\omega)=i[\Sigma_{P}^{R}(\omega)-\Sigma_{P}^{A}(\omega)]$. The time-averaging steps that reduce the double-time current formula to this stationary-contact expression are summarized in Appendix~\ref{sec:Javg}. In the weak-probe limit $\mathbf{\tilde{G}}\simeq\mathbf{\tilde{g}}$, so the voltage dependence is controlled by the time-averaged local spectral function
\begin{equation}\label{eq:Aspectroscopy}
	A_{\mathrm{loc}}^{\mathrm{ave}}(\omega)=-\frac{1}{\pi}\mathrm{Im}\,\mathrm{tr}\,\mathbf{\tilde{g}}_{\mathrm{loc}}^{R}(\omega,0).
\end{equation}
The nonequilibrium occupation of the sliding CDW enters through $\mathbf{\tilde{g}}^{<}$, whereas the singular structures in $d\langle J_{P}\rangle/dV_{P}$ are governed by Eq.~\eqref{eq:Aspectroscopy}. Thus weak-contact tunneling spectroscopy directly probes the split gap edges and Floquet sidebands generated by the sliding CDW.

\subsection{Near-edge model and quantitative comparison with experiment}

To connect most directly with experiment, we evaluate a near-edge model in which Eq.~\eqref{eq:Aspectroscopy} is represented by a thermally broadened, Dynes-regularized split-edge envelope in a narrow window around the first positive gap edge. The sliding CDW sustains a macroscopic longitudinal voltage drop, so the local electrostatic potential acts as a rigid reference for the local CDW bands. The applied probe voltage $V_{P}$ sets the probe chemical potential $\mu_{P}$ relative to this local environment. For the (TaSe$_4$)$_2$I nanowire data of Ref.~\cite{Le24}, the measured device voltage varies only weakly with current once the system is above depinning. As a first approximation, we therefore treat the local probe offset $\epsilon=\mu_{P}-E_{v}$ (where $E_{v}$ is the local lower gap edge) as slowly varying while $\Omega$ continues to increase linearly with the imposed through-current via Eq.~\eqref{eq:Omega-I}. Appendix~\ref{sec:realistic-single-contact} gives the explicit formulas for the broadened single-contact spectra and the inverse-current oscillatory contribution.

\subsection{Percolating coherent filament: the central physical deduction}

Quantitative matching between theory and experiment yields one of the central physical insights of this work. Using the experimental transport gap $\Delta=165\,\mathrm{meV}$ and oscillation period $\Delta(1/I)=0.00269\,\mu\mathrm{A}^{-1}$ from Ref.~\cite{Le24}, a representative contact mismatch $\epsilon=10\,\mathrm{meV}$ implies an effective coherent channel number $N_{\mathrm{eff}}\simeq4.8\times10^{2}$ and $\hbar\Omega\simeq0.54\,\mathrm{meV}$ at $I=20\,\mu\mathrm{A}$.

\begin{figure}[tb]
	\centering
	\resizebox{\linewidth}{!}{%
		\begin{tikzpicture}[
			x=1cm,y=1cm,
			bulk/.style={draw=black, thick, fill=gray!15, rounded corners=2pt},
			contact/.style={draw=black, thick, fill=orange!30, rounded corners=1pt},
			filament/.style={line width=6pt, cyan!30, smooth, tension=0.6},
			cdwline/.style={thick, blue!70!black, smooth, tension=0.6},
			lab/.style={font=\small, align=center},
			curr/.style={very thick, -{Latex[length=2.2mm]}, red!80!black}
			]
			\path[bulk] (0,0) rectangle (7, 2.2);
			\node[lab, text=gray!80!black] at (3.5, 1.8) {Pinned Bulk ($N_{\mathrm{geom}} \sim 3 \times 10^4$)};
			\draw[filament] plot coordinates {(-0.1, 0.7) (1.5, 1.2) (3.5, 0.6) (5.5, 1.1) (7.1, 0.7)};
			\draw[cdwline] plot coordinates {(-0.1, 0.7) (1.5, 1.2) (3.5, 0.6) (5.5, 1.1) (7.1, 0.7)};
			\node[lab, text=blue!70!black] at (3.5, 0.2) {Coherent Sliding Filament ($N_{\mathrm{eff}} \ll N_{\mathrm{geom}}$)};
			\draw[thick, gray!60, dashed, -Latex] (0, 0.7) -- (1.5, 1.2);
			\draw[thick, gray!60, dashed, -Latex] (7, 0.7) -- (5.5, 1.1);
			\path[contact] (-0.6, 0.3) rectangle (0.2, 1.1);
			\path[contact] (6.8, 0.3) rectangle (7.6, 1.1);
			\node[lab] at (-0.2, 0.7) {$I_{\mathrm{in}}$};
			\node[lab] at (7.2, 0.7) {$I_{\mathrm{out}}$};
			\draw[curr] (-1.6, 0.7) -- node[above, text=red!80!black, font=\small, inner sep=2pt] {$I$} (-0.7, 0.7);
			\draw[curr] (7.7, 0.7) -- node[above, text=red!80!black, font=\small, inner sep=2pt] {$I$} (8.6, 0.7);
		\end{tikzpicture}
	}
	\caption{\label{Fig:filament} Percolating sliding filament deduced from quantitative matching between theory and experiment. The macroscopic terminal-to-terminal current $I$ must funnel through a highly localized active region, with effective coherent channel number $N_{\mathrm{eff}}\sim 480$, nearly two orders of magnitude smaller than the geometric chain count $N_{\mathrm{geom}}\sim 3\times 10^{4}$ of the pinned bulk. This filamentary confinement is essential for observability (see text).}
\end{figure}
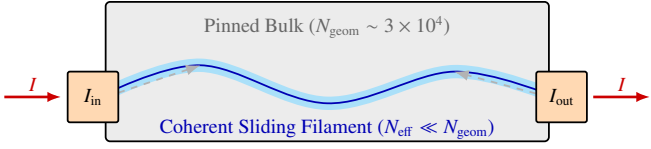

Strikingly, typical device dimensions~\cite{Le24} yield a geometric chain count $N_{\mathrm{geom}} \sim 3 \times 10^4$, nearly two orders of magnitude larger than the deduced $N_{\mathrm{eff}}$. This inequality $N_{\mathrm{eff}} \ll N_{\mathrm{geom}}$ is a profound physical signature: it dictates that the sliding state does not depin as a monolithic bulk. Instead, as sketched in Fig.~\ref{Fig:filament}, the macroscopic terminal-to-terminal current $I$ must percolate through a narrow, highly coherent active filament.

Several physical mechanisms can produce such filamentary confinement. First, surface-dominated tunneling may cause the metallic probe to couple preferentially to the outermost CDW chains, effectively reducing the cross-section sampled by the spectroscopy. Second, the CDW may depin filamentarily through transverse domains, with collective sliding establishing first along a preferred path before propagating laterally. Third, a finite transverse phase-correlation length may limit the number of chains that participate coherently in the local Floquet spectroscopy, even when the underlying depinning is more uniform. Distinguishing among these mechanisms requires spatially resolved probes beyond the scope of the present theory, but all of them lead to the same effective description $N_{\mathrm{eff}} \ll N_{\mathrm{geom}}$.

Crucially, this filamentary confinement is precisely what makes the observation of these quantum oscillations experimentally feasible. Achieving the required sliding frequency $\Omega$ uniformly across the entire bulk $N_{\mathrm{geom}}$ would require a total applied current nearly two orders of magnitude larger; the resulting macroscopic Joule heating would thermally destroy the delicate Floquet phase coherence. Indeed, if one inserts the full geometric chain count, $\hbar\Omega$ falls far below the thermal and Dynes broadening widths, so no resolvable sideband structure could exist. The observed spectroscopy therefore must be controlled by the much smaller contact-local effective channel number identified above.

\subsection{Fixed-bias inverse-current oscillations}

\begin{figure*}[tb]
\centering
\includegraphics[width=\linewidth]{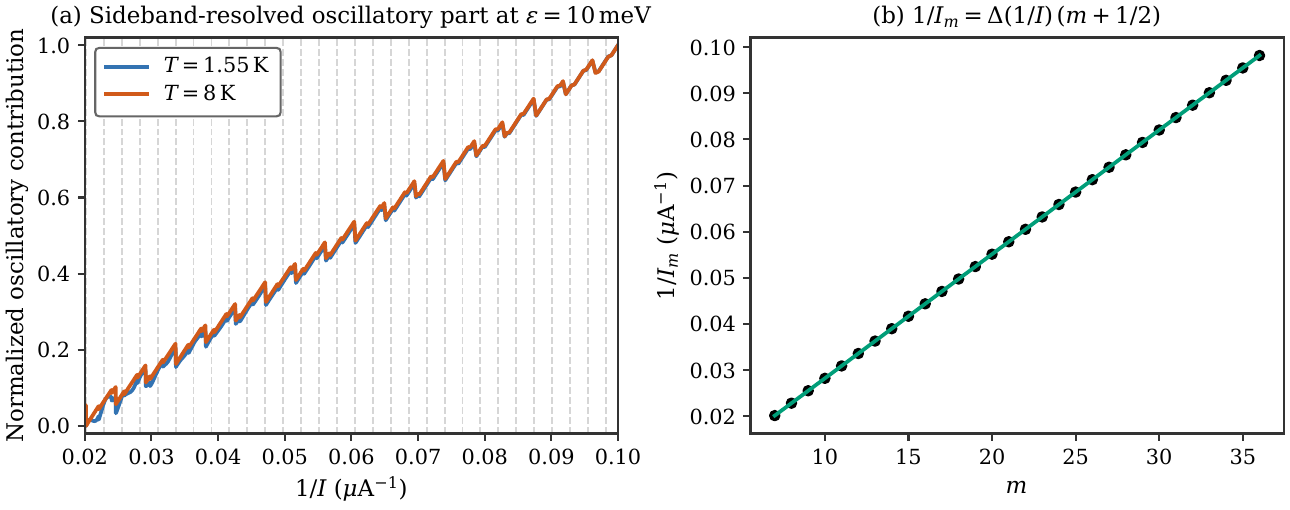}
\caption{\label{Fig:inverseI} Fixed-bias inverse-current oscillations generated by the Floquet sideband ladder. (a) Individually normalized oscillatory contribution $\delta G_{\mathrm{osc}}(I,\epsilon;T)$ at the reference offset $\epsilon=10\,\mathrm{meV}$ as a function of $1/I$. (b) Corresponding crossing positions, which obey $1/I_{m}=\Delta(1/I)(m+\frac{1}{2})$ with $\Delta(1/I)=0.00269\,\mu\mathrm{A}^{-1}$, matching the experimental period of Ref.~\cite{Le24}. The oscillatory contribution $\delta G_{\mathrm{osc}}$ is evaluated from Eq.~\eqref{eq:fixedBiasOscillation} of Appendix~\ref{sec:realistic-single-contact}.}
\end{figure*}

At fixed bias offset $\epsilon=eV-E_{v}$, successive sideband edges of the Floquet ladder cross the probe chemical potential whenever
\begin{equation}\label{eq:resonance}
\epsilon\simeq\left(m+\frac{1}{2}\right)\hbar\Omega(I_{m}),\qquad m=0,1,2,\ldots,
\end{equation}
so that $1/I_{m}$ is linear in $m$. This mechanism provides a striking temporal analog of the Shubnikov--de Haas effect. In conventional magnetic quantum oscillations, periodicity in $1/B$ arises because Landau levels with energy spacing $\hbar\omega_{c} \propto B$ successively cross a fixed Fermi energy. Here, periodicity in $1/I$ arises because Floquet sidebands with energy spacing $\hbar\Omega \propto I$ successively cross a fixed contact mismatch energy $\epsilon$. Figure~\ref{Fig:inverseI} shows the resulting fixed-bias inverse-current oscillation evaluated from the near-edge model of Appendix~\ref{sec:realistic-single-contact}.

\begin{figure*}[tb]
\centering
\includegraphics[width=\textwidth]{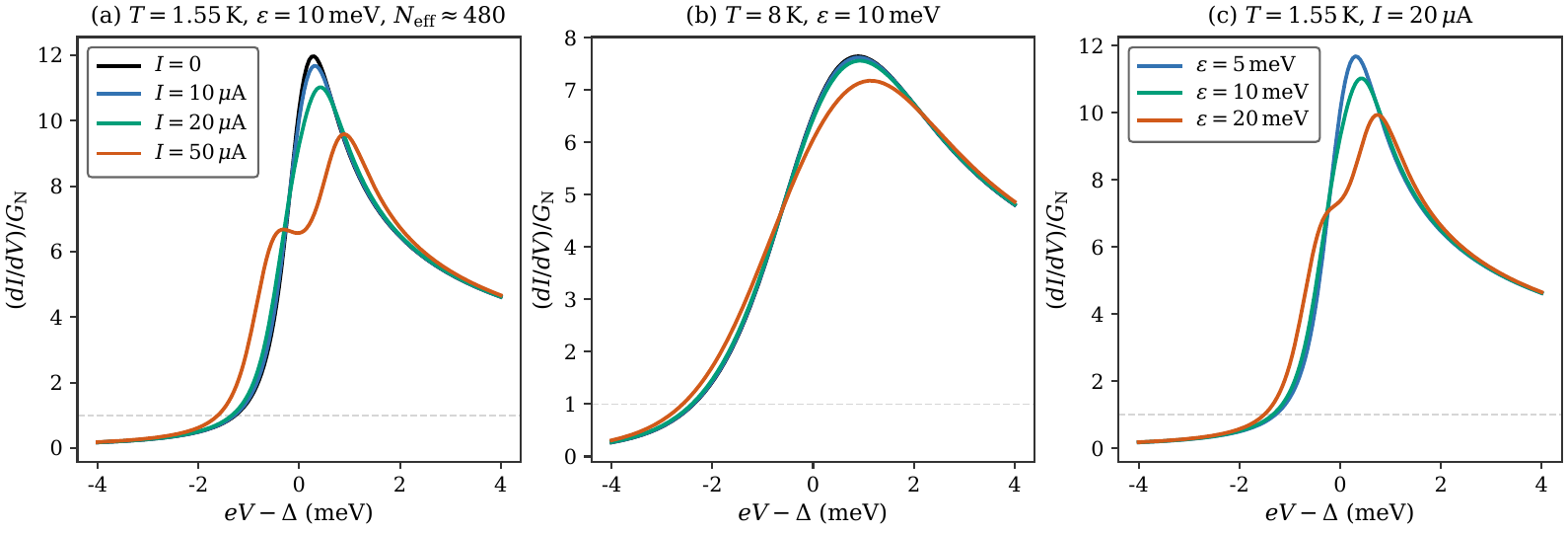}
\caption{\label{Fig:tunnelSpec} Near-edge model calculation of single-contact tunneling spectra near the positive CDW gap edge, obtained by thermally convolving a Dynes-regularized split-edge envelope motivated by Eq.~\eqref{eq:Aspectroscopy} with broadening $\gamma=0.30\,\mathrm{meV}$. (a) Reference low-temperature spectra at $T=1.55\,\mathrm{K}$ for $\epsilon=10\,\mathrm{meV}$, corresponding to $N_{\mathrm{eff}}\simeq4.8\times10^{2}$. (b) Same parameter set at $T=8\,\mathrm{K}$. (c) Sensitivity at $T=1.55\,\mathrm{K}$ and $I=20\,\mu\mathrm{A}$ for $\epsilon=5,10,20\,\mathrm{meV}$, corresponding to $N_{\mathrm{eff}}\simeq960,480,240$. Only the positive-bias edge is shown.}
\end{figure*}

The same near-edge construction also yields the fixed-current cuts of $G(V,I)$ near the singular edge structure shown in Fig.~\ref{Fig:tunnelSpec}. The single-contact spectra and the inverse-current oscillation are therefore complementary cuts of the same Floquet sideband ladder. Reference~\cite{Le24} pointed out the basic periodicity relation $\epsilon = (m+\frac{1}{2})\hbar\Omega(I_m)$; the present analysis additionally shows that the same ladder controls the full near-edge spectroscopy at fixed current, providing a unified picture in the $(V,I)$ plane and revealing the physical origin of the coherent filament through quantitative parameter extraction.

\section{Transport through contacted sliding-CDW devices}
\label{sec:transport}

Section~\ref{sec:spectroscopy} established that a weak single contact directly probes the local Floquet spectrum. The experiment of Ref.~\cite{Le24}, however, employs a persistent current source and measures terminal-pair voltages across multiple voltage probes simultaneously. The macroscopic depinned sliding state must therefore be analyzed within a genuine multiterminal setting, where one common through-current coexists with terminal-dependent oscillation visibility. The principal puzzle is that the inner-terminal voltage $V_{2-3}^{1-4}(I)$ exhibits a clean $1/I$ oscillation, whereas the raw outer-terminal voltage $V_{1-4}^{1-4}(I)$ reveals the same kernel only after smooth-background subtraction. Reference~\cite{Le24} reported this visibility hierarchy but did not provide a transport-level interpretation. We now develop such an interpretation through a minimal segmented multiterminal model (Sec.~\ref{sec:segmented-main}), and we contrast it with a homogeneous two-terminal Floquet-Landauer reference limit (Sec.~\ref{sec:twoTerminal-main} and Appendix~\ref{sec:near-threshold-two-terminal}) that helps identify which features depend on spatial inhomogeneity.

\subsection{Current-driven segmented multiterminal model}
\label{sec:segmented-main}

The directly relevant experimental observable is a current-driven terminal-pair voltage $V_{p-q}^{1-4}(I)$ in a multiterminal geometry~\cite{Buttiker1986}, rather than the differential conductance of a spatially uniform voltage-biased span. Because Ref.~\cite{Le24} uses a persistent current source, the steady-state control variable is the common through-current and the terminal voltages are measured responses; throughout this section ``current driven'' is used in this sense. Motivated by this geometry, we introduce a minimal segmented description in which the current path is decomposed into three series segments $1$-$2$, $2$-$3$, and $3$-$4$, all carrying the same imposed current. The coherent Floquet oscillatory kernel is assumed to be generated mainly in the inner segment $2$-$3$ of the active percolating filament identified in Sec.~\ref{sec:spectroscopy}. The measurement logic was already sketched in Fig.~\ref{Fig:segmentedSchematic}.

\begin{figure*}[tb]
	\centering
	\includegraphics[width=\textwidth]{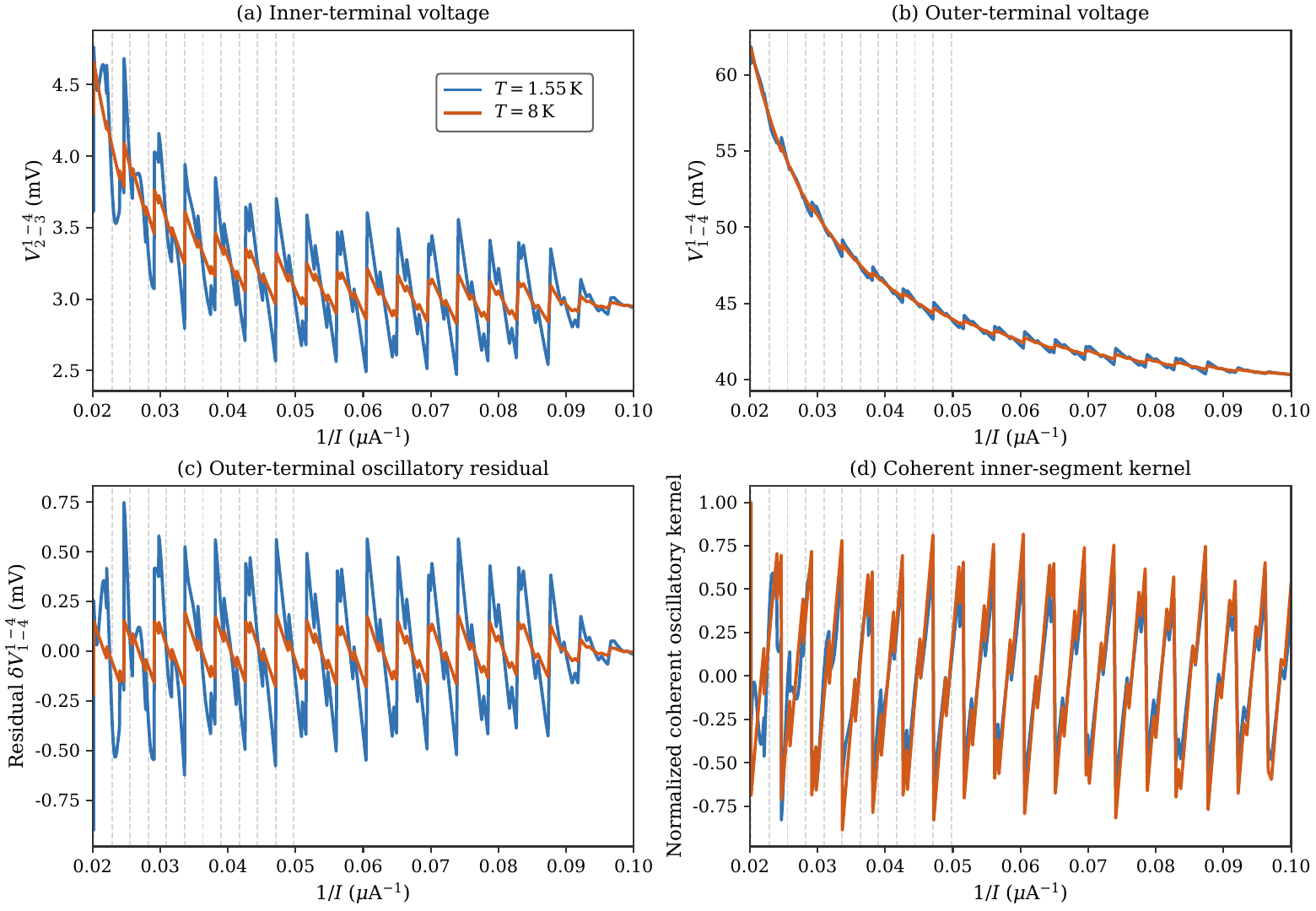}
	\caption{\label{Fig:segmentedTransport} Minimal current-driven segmented model for the experimental measurement geometry of Ref.~\cite{Le24} with current driven between terminals $1$ and $4$. (a) Inner-terminal voltage $V_{2-3}^{1-4}(I)$ versus $1/I$. (b) Raw outer-terminal voltage $V_{1-4}^{1-4}(I)$ versus $1/I$. (c) Oscillatory residual obtained after subtracting the smooth background from panel (b). (d) Normalized coherent kernel $\mathcal{O}_{\mathrm{osc}}(I;T)$. In this minimal visibility model the coherent response is concentrated in segment $2$-$3$, while segments $1$-$2$ and $3$-$4$ contribute mainly smooth series voltages due to phase-slip dephasing near the contacts. Explicit definitions are given by Eqs.~\eqref{eq:segmentedKernel}--\eqref{eq:segmentedMeasured} of Appendix~\ref{sec:segmented-transport}.}
\end{figure*}

In this minimal version, the outer segments are effectively dephased and contribute only smooth series voltages. Physically, this loss of coherence is a direct consequence of the finite phase-coherence length of the device combined with strong inelastic scattering near the current-injecting contacts. To convert the collective sliding CDW current into a normal electron current at the outer metallic leads, dissipative phase-slip processes must occur, locally destroying the uniform Floquet phase coherence~\cite{LarkinVarlamov2009}. Consequently, on length scales larger than the phase-coherence length, the active filament behaves as a classical series combination of coherent and incoherent subregions~\cite{Imry2002}, with the strongly dephased outer segments yielding only smooth macroscopic voltage drops. Then $V_{2-3}^{1-4}(I)$ retains a clear $1/I$ oscillation, whereas $V_{1-4}^{1-4}(I)$ is dominated by smooth background drops and reveals the same kernel only after background subtraction. Figure~\ref{Fig:segmentedTransport} shows this visibility effect explicitly. Appendix~\ref{sec:segmented-transport} gives the explicit effective construction.

This effective construction is not intended to reproduce the full nonlinear waveform of the experiment nor the precise amplitude. Rather, it isolates how a single common current can coexist with strongly terminal-dependent oscillation visibility through the interplay of coherent (inner) and dephased (outer) segments. The qualitative agreement with the experimentally observed hierarchy supports the underlying physical picture: phase-slip-induced dephasing localized near the contacts, combined with a coherent filament across the inner span, naturally produces the observed visibility pattern.

\subsection{Homogeneous two-terminal reference limit}
\label{sec:twoTerminal-main}

For orientation it is useful to compare the segmented multiterminal model with a homogeneous two-terminal reference problem sketched in Fig.~\ref{Fig:device}. Consider a finite sliding-CDW region $H_{D}(t)$ coupled to left and right reservoirs. In contrast to the weak-probe setup of Sec.~\ref{sec:spectroscopy} and the segmented construction above, here the dc current and the nonequilibrium distribution are determined by the contacts themselves. In the physical electron basis the lead self-energies are stationary, while the periodicity resides in the isolated device Green's function $\mathbf{\tilde{g}}$. For a noninteracting device in the periodic steady state~\cite{Jauho94,Stefanucci2013,HaugJauho2008}, the contacted Green's functions satisfy
\begin{equation}\label{eq:DysonTwoTerminal}
	\mathbf{\tilde{G}}^{R/A}=\mathbf{\tilde{g}}^{R/A}+\mathbf{\tilde{g}}^{R/A}\left(\Sigma_{L}^{R/A}+\Sigma_{R}^{R/A}\right)\mathbf{\tilde{G}}^{R/A},
\end{equation}
\begin{equation}\label{eq:LessTwoTerminal}
	\mathbf{\tilde{G}}^{<}=\mathbf{\tilde{G}}^{R}\left(\Sigma_{L}^{<}+\Sigma_{R}^{<}\right)\mathbf{\tilde{G}}^{A},
\end{equation}
where products denote matrix multiplication in the device space together with the time convolutions appropriate for the double-time Green's functions. Appendix~\ref{sec:sk-dyson} summarizes the corresponding Schwinger-Keldysh Dyson structure in the physical electron basis. The time-averaged current through the left contact is
\begin{equation}\label{eq:JLavg}
	\langle J_{L}\rangle = -\frac{e}{\hbar}\mathrm{Im}\int\frac{d\omega}{2\pi}\mathrm{tr}\left\{\mathbf{\Gamma}^{L}(\omega)\left[\mathbf{\tilde{G}}^{<}(\omega,0)+2f_{L}(\omega)\mathbf{\tilde{G}}^{R}(\omega,0)\right]\right\},
\end{equation}
with an analogous expression for $\langle J_{R}\rangle$ and charge conservation $\langle J_{L}\rangle=-\langle J_{R}\rangle$ in the steady state.

These equations define the homogeneous periodically driven two-terminal problem: the sliding CDW contributes sideband-assisted transmission channels with exchanged energies $n\hbar\Omega$, while the nonequilibrium distribution is fixed by the two reservoirs. A fully self-consistent nonlinear $I$-$V$ curve would additionally require a microscopic relation between the through-current, the depinned sliding frequency, and the spatial distribution of contact voltage drops. As a reference transport calculation, we therefore keep $\hbar\Omega$ fixed and evaluate the near-threshold differential conductance around $eV\simeq2\Delta$, where the response is controlled by the singular Floquet gap edges. The inward sideband thresholds occur at
\begin{equation}\label{eq:twoTerminalThresholdMain}
	eV_{m}\simeq2\Delta-(2m+1)\hbar\Omega,\qquad m=0,1,2,\ldots,
\end{equation}
so the response shifts linearly with $\hbar\Omega$.

\begin{figure*}[tb]
	\centering
	\includegraphics[width=\textwidth]{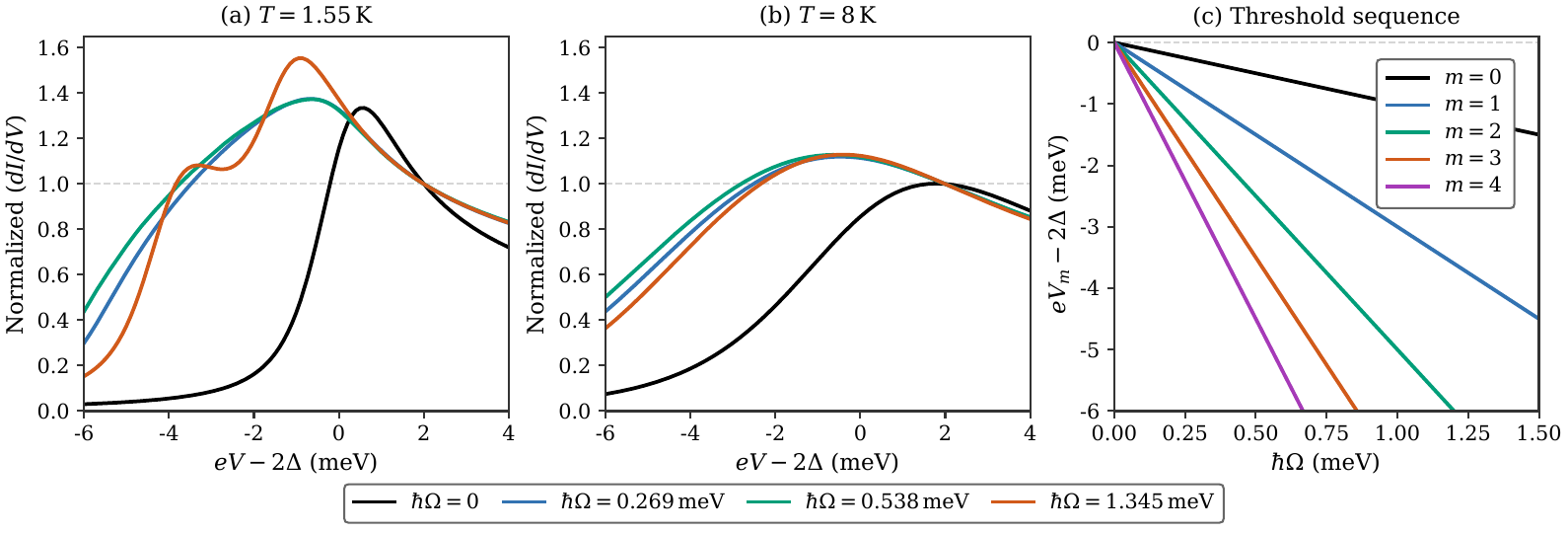}
	\caption{\label{Fig:twoTerminal} Homogeneous voltage-biased reference calculation for a uniformly coherent sliding span at fixed $\hbar\Omega$. (a,b) Normalized $dI/dV$ traces as functions of $eV-2\Delta$ for $\hbar\Omega=0$, $0.269$, $0.538$, and $1.345\,\mathrm{meV}$, at $T=1.55\,\mathrm{K}$ and $8\,\mathrm{K}$, respectively. Each curve is normalized by its value at $eV-2\Delta=2\,\mathrm{meV}$ to emphasize the threshold shifts. (c) Threshold sequence $eV_{m}=2\Delta-(2m+1)\hbar\Omega$.}
\end{figure*}

Figure~\ref{Fig:twoTerminal} shows the resulting fixed-$\hbar\Omega$ near-threshold conductance for the same realistic gap and broadening scales used in the single-contact estimate. This homogeneous calculation identifies the threshold structure expected from a uniformly coherent sliding span under voltage bias, but it is not the persistent-current-driven multiterminal experiment of Ref.~\cite{Le24}. In particular, it tracks threshold motion at fixed $\hbar\Omega$ rather than the fixed-offset sideband crossings that generate the observed $1/I$ sequence. The homogeneous two-terminal limit thus serves as a useful theoretical reference for identifying universal features of the Floquet sideband ladder under different bias conditions, but the experimentally relevant signal requires the segmented current-driven analysis of Sec.~\ref{sec:segmented-main}. Appendix~\ref{sec:near-threshold-two-terminal} gives the explicit near-threshold model used to evaluate Fig.~\ref{Fig:twoTerminal}.
	
\section{Discussion and outlook}
\label{sec:discussion}

The exact sliding-CDW Floquet solution of Sec.~\ref{sec:floquet} generates a single sideband ladder, but different measurements project that ladder onto different observables. A weak single contact reads it as local spectroscopy (Sec.~\ref{sec:spectroscopy}); once the depinned state is entered and the local contact offset varies only weakly with current, the same ladder produces a fixed-bias $1/I$ sequence through repeated sideband crossings. A persistent-current-driven multiterminal device reads it through terminal-dependent voltage drops (Sec.~\ref{sec:segmented-main}), whereas a homogeneous voltage-biased span reads it as threshold shifts in the $(V,\Omega)$ plane (Sec.~\ref{sec:twoTerminal-main}). The experiment of Ref.~\cite{Le24} belongs to the multiterminal category, so spatial inhomogeneity and contact dephasing are essential for understanding why the inner-terminal signal exhibits a much stronger $1/I$ oscillation than the raw outer-terminal voltage.

\subsection{Physical implications of the percolating filament}

A central physical deduction of our framework is the true nature of the macroscopically observed sliding state. The parameter $N_{\mathrm{eff}}$ represents the effective coherent channel number of a percolating filament, not a literal geometric chain count of the whole nanowire. For typical device dimensions~\cite{Le24}, the geometric chain count $N_{\mathrm{geom}} \sim 3 \times 10^4$ is nearly two orders of magnitude larger than the $N_{\mathrm{eff}} \sim 480$ required to reproduce the experimental oscillation period at a reasonable bias offset. This strong reduction $N_{\mathrm{eff}} \ll N_{\mathrm{geom}}$ implies that the macroscopic quantum coherence probed by the contacts is highly spatially localized.

This filamentary picture leads to several testable predictions for future experiments. First, $N_{\mathrm{eff}}$ should scale weakly (if at all) with the geometric cross-section of nanowires above some threshold, since coherent transport is confined to a narrow active path rather than spanning the full bulk. Second, deliberate manipulation of the contact geometry, surface treatment, or disorder landscape should modify $N_{\mathrm{eff}}$ and shift the oscillation period accordingly. Third, spatially resolved probes such as scanning tunneling microscopy or microwave near-field imaging should reveal localized coherent regions during the sliding state. Fourth, the predicted absence of $1/I$ oscillations in two-dimensional CDWs with residual Fermi surfaces, due to severe electron-electron dephasing of the Floquet sidebands, should be testable in comparative studies across different CDW materials.

The effective spatial confinement is also a vital protective mechanism. By reaching high sliding frequencies with a relatively small total applied current, the active region avoids the severe macroscopic Joule heating that would otherwise induce rapid thermal dephasing. In this sense, the filamentary confinement is not incidental but a precondition for the observation itself: it is precisely the reduction $N_{\mathrm{eff}} \ll N_{\mathrm{geom}}$ that allows the experimentally accessed current range to produce Floquet frequencies above the thermal and Dynes broadening scales.

\subsection{Open puzzles and limitations}

Several experimental features of Ref.~\cite{Le24} remain outside the present treatment and deserve future attention. First, the oscillation re-entrance at elevated temperatures observed in device D3 (Appendix G of Ref.~\cite{Le24}), where a second oscillation sequence with a different period appears above the suppression of the low-temperature sequence, is not captured by the present framework. This re-entrance suggests temperature-driven reorganization of the active filament or of the effective contact offset $\epsilon$, both of which would require a more detailed microscopic model of the depinning dynamics. Second, our theory predicts a phase shift $b=1/2$ for the linear $1/I_m$ vs $m$ relation, while the experimental fits report device-dependent values $b$ ranging from approximately $-0.16$ to $0.30$, with deviations more pronounced at high currents where heating effects also distort the periodicity. Reconciling this systematic discrepancy likely requires accounting for sideband-dependent matrix elements, broadened edge profiles, or weak current-induced renormalization of the contact offset. Third, the slight shifts of oscillation positions in transverse magnetic field observed on certain devices remain unexplained within the present framework.

Beyond these specific puzzles, our segmented current-driven model is intended as a minimal transport interpretation of the visibility hierarchy along the filament, not as a quantitative fit to the oscillation amplitude or to the full sawtooth waveform of the experiment. A future material-specific microscopic theory would have to self-consistently determine how the imposed persistent current sets the local voltage drops, how phase-slip conversion near the contacts~\cite{LarkinVarlamov2009} feeds back on the sliding frequency, and how additional non-CDW transport channels modify the measured terminal voltages. The present framework provides a transport-level scaffold on which such material-specific microscopic theories can be built.

\subsection{Broader implications}

Within these limitations, the sequence of exact Floquet solution, weak-probe spectroscopy, segmented current-driven modeling, and homogeneous transport reference provides a coherent theoretical framework for the observed inverse-current oscillations. More broadly, the universal spatial-to-temporal conversion identified here establishes a critical material prerequisite for observing macroscopic sliding coherence. We emphasize that these $1/I$ quantum oscillations are expected to emerge only in fully gapped CDW insulators. In conventional two-dimensional CDWs possessing a residual Fermi surface, low-energy electron-electron scattering induces rapid dephasing that quickly smears out the delicate Floquet sidebands. By contrast, the complete excitation gap of a quasi-1D CDW insulator effectively freezes out these scattering channels, protecting the coherent Floquet dynamics.

This intrinsic dc-to-ac conversion mechanism offers a compelling design principle for future quantum devices. Highly coherent, tunable high-frequency drives can in principle be generated locally via simple dc currents, without the need for external microwave architecture. The same universal mechanism may apply to other sliding ordered states, including spin-density waves and certain density-wave phases of strongly correlated electron systems. In each case, the basic prerequisite remains the same: a coherent sliding mode protected from low-energy dephasing by an excitation gap. Identifying and engineering such gap-protected coherent sliding states across material platforms is a natural direction for future work.
	
\begin{acknowledgments}
	The author thanks T.~Le, F.~M.~Qu, and L.~Lu for stimulating discussions on the experimental phenomenology of Ref.~\cite{Le24} that motivated this theoretical investigation. This work was supported by the National Key R\&D Program of China (Grant No. 2022YFA1403403) and the National Natural Science Foundation of China (Grant Nos. 12274441 and 12534004).
\end{acknowledgments}
	
\section*{Data and Code Availability}

To ensure full transparency and reproducibility, all materials associated with this study have been made publicly available in the GitHub repository \texttt{Sliding-CDW}~\cite{yizhou76-sudo2026SlidingCDW}. The repository contains:
\begin{itemize}
	\item \textbf{AI-assisted research transcripts:} Complete, unedited transcripts of conversations with large language models employed during manuscript preparation, including interactions with Gemini 3.1 Pro Preview (Google), GPT-5.4 (OpenAI), Claude 4.6 Opus (Anthropic), and the AI agent Codex (OpenAI). These tools were used for code generation, manuscript polishing, and consistency checking; all physical content, derivations, and conclusions were developed and verified by the author.
	\item \textbf{Computational scripts:} Python scripts used to generate all numerical results and figures presented in this work, together with documentation and dependency specifications.
\end{itemize}

\appendix
	
\section{Exact solution of a general two-level Floquet system}\label{sec:2L}

The exact diagonalization used in Sec.~\ref{sec:floquet} is a special case of the following general result. Consider the time-dependent two-level Hamiltonian
\begin{equation}\label{eq:H2L}
	H=\left[\begin{array}{cc}a&b e^{-i(\phi_0-\Omega t)}\\b e^{i(\phi_0-\Omega t)}&-a\end{array}\right],
\end{equation}
with real parameters $a$, $b$, $\Omega$, and $\phi_0$. The Floquet eigenvalue problem reads
\begin{equation}\label{eq:SH2L}
	\left(H-i\frac{\partial}{\partial t}\right)\left(\begin{array}{c}u\\v\end{array}\right)=E\left(\begin{array}{c}u\\v\end{array}\right).
\end{equation}
Expanding $u(t)=\sum_{n}u_{n}e^{-in\Omega t}$ and $v(t)=\sum_{n}v_{n}e^{-in\Omega t}$, each pair $(u_{n},v_{n+1})$ satisfies
\begin{equation}
	\left[\begin{array}{cc} a+n\Omega-E & b e^{-i\phi_0}\\ b e^{i\phi_0} & -a+(n+1)\Omega-E \end{array}\right]\left(\begin{array}{c}u_{n}\\v_{n+1}\end{array}\right)=0.
\end{equation}
The problem closes in each $(u_{n},v_{n+1})$ sector and yields the quasienergies
\begin{equation}
	E_{n\pm}=\left(n+\frac{1}{2}\right)\Omega \pm \mathcal{E},\qquad \mathcal{E}=\sqrt{\left(a-\frac{\Omega}{2}\right)^2+b^2}.
\end{equation}
The corresponding eigenvectors diagonalize the static effective matrix
\begin{equation}
	\left[\begin{array}{cc} a-\frac{\Omega}{2} & b e^{-i\phi_0}\\ b e^{i\phi_0} & -a+\frac{\Omega}{2} \end{array}\right],
\end{equation}
and are independent of both $t$ and the sideband index $n$:
\begin{equation}\label{eq:uv}
	\left(\begin{array}{c}u_{n}\\v_{n+1}\end{array}\right)_{+} = \left[\begin{array}{c}\cos\frac{\theta}{2}e^{-i\phi_{0}}\\ \sin\frac{\theta}{2}\end{array}\right],\quad \left(\begin{array}{c}u_{n}\\v_{n+1}\end{array}\right)_{-} = \left[\begin{array}{c}-\sin\frac{\theta}{2}e^{-i\phi_{0}}\\\cos\frac{\theta}{2}\end{array}\right]
\end{equation}
with $\cos\theta=(a-\Omega/2)/\mathcal{E}$. Defining
\begin{equation}
	S_{n}(t) = \left[\begin{array}{cc} e^{-in\Omega t} & 0 \\ 0 & e^{-i(n+1)\Omega t} \end{array}\right]\left[\begin{array}{cc} u_{n,+} & u_{n,-} \\ v_{n+1,+} & v_{n+1,-} \end{array}\right],
\end{equation}
one verifies that
\begin{equation}
	\left(H-i\frac{\partial}{\partial t}\right)S_{n}=S_{n}\left[\mathcal{E}\sigma_3+\left(n+\frac{1}{2}\right)\Omega\right],
\end{equation}
so that the Floquet Hamiltonian is diagonal in the $S_n$ basis. This result, with the identifications $a \leftrightarrow (\varepsilon_{\mathbf{k}}-\varepsilon_{\mathbf{k}+r\mathbf{Q}})/2$ and $b \leftrightarrow \Delta$, underlies the sliding-CDW spectrum given by Eq.~\eqref{eq:Ek} of the main text.
	
\section[Integral Floquet spectral functions $A_{n}(E)$]{Integral Floquet spectral functions \texorpdfstring{$A_{n}(E)$}{An(E)}}\label{sec:An}

This appendix derives the time-averaged Floquet density of states quoted in Sec.~\ref{sec:floquet}. Near the Fermi level, $\xi_{\mathbf{k}}=\varepsilon_{\mathbf{k}}-\varepsilon_{F}\simeq\hbar v_{F}(|\mathbf{k}|-k_{F})$ and the normal-state DOS is $N(\xi)\simeq N(0)=1/(\pi\hbar v_{F})$. Then
\begin{equation}
	\mathcal{E}_{\mathbf{k}}= \sqrt{\left(\xi_{\mathbf{k}}-\frac{r\hbar\Omega}{2}\right)^2+\Delta^2}.
\end{equation}
Because $E_{\mathbf{k}n}=E_{\mathbf{k}0}+n\hbar\Omega$, the sideband-resolved spectral functions satisfy
\begin{equation}
	A_{n}(E) \equiv-\frac{1}{\pi}\mathrm{Im}\sum_{\mathbf{k}}g_{\mathbf{k}}^{R}(E,n) = A_{0}(E-n\hbar\Omega).
\end{equation}
Evaluating the $n=0$ component gives
\begin{widetext}
	\begin{equation}
		\begin{split}
			A_{0}(E) &= \sum_{\mathbf{k}}\delta\left(E-E_{\mathbf{k}0}\right) = N(\xi_{\mathbf{k}})\left|\frac{dE_{\mathbf{k}0}}{d\xi_{\mathbf{k}}}\right|^{-1}_{E_{\mathbf{k}0}=E} \\
			&= \frac{N(0)\left|E-\varepsilon_{F}+\frac{1}{2}\hbar\Omega\right|}{2\sqrt{\left(E-\varepsilon_{F}+\frac{1}{2}\hbar\Omega\right)^2-\Delta^2}}\theta\left(|E-\varepsilon_{F}+\tfrac{1}{2}\hbar\Omega|-\Delta\right) + \frac{N(0)\left|E-\varepsilon_{F}-\frac{1}{2}\hbar\Omega\right|}{2\sqrt{\left(E-\varepsilon_{F}-\frac{1}{2}\hbar\Omega\right)^2-\Delta^2}}\theta\left(|E-\varepsilon_{F}-\tfrac{1}{2}\hbar\Omega|-\Delta\right).
		\end{split}
	\end{equation}
\end{widetext}
Defining the static DOS by $D(E)=A_{0}(E)|_{\Omega=0}$, we obtain
\begin{equation}
	D(E) = \frac{N(0)\left|E-\varepsilon_{F}\right|}{\sqrt{\left(E-\varepsilon_{F}\right)^2-\Delta^2}}\theta\left(|E-\varepsilon_{F}|-\Delta\right),
\end{equation}
and
\begin{equation}
	A_{0}(E)=\frac{1}{2}\left[D\left(E+\frac{\hbar\Omega}{2}\right)+D\left(E-\frac{\hbar\Omega}{2}\right)\right].
\end{equation}
This is the split-edge structure illustrated in Fig.~\ref{Fig:DOS}.

\section{Basis transformation between Floquet quasiparticles and physical electrons}\label{sec:noninteracting-basis}

This appendix gives the explicit transformation between the diagonal Floquet quasiparticle basis of Sec.~\ref{sec:floquet} and the physical electron basis in which the lead self-energies (Appendix~\ref{sec:lead-self-energy}) are stationary. The quasiparticle operators $\{d_{\mathbf{k}\sigma}^{\dagger}\}$ and the physical electron operators $\{f_{\mathbf{k}\sigma}^{\dagger}\}$ are related by the time-dependent unitary matrix
\begin{equation}
	U_{\mathbf{k}}(t)=\left(\begin{array}{cc} u_{\mathbf{k}} & v_{\mathbf{k}}e^{-ir\Omega t} \\ v_{\mathbf{k}+r\mathbf{Q}}e^{ir\Omega t} & u_{\mathbf{k}+r\mathbf{Q}} \end{array}\right),
\end{equation}
so that
\begin{equation}
	\begin{aligned}
		\left(\begin{array}{c} d_{\mathbf{k}\sigma}^{\dagger}\\ d_{\mathbf{k}+r\mathbf{Q},\sigma}^{\dagger} \end{array}\right) &=U_{\mathbf{k}}(t)\left(\begin{array}{c} f_{\mathbf{k}\sigma}^{\dagger}\\ f_{\mathbf{k}+r\mathbf{Q},\sigma}^{\dagger} \end{array}\right),\\
		\left(\begin{array}{c} f_{\mathbf{k}\sigma}\\ f_{\mathbf{k}+r\mathbf{Q},\sigma} \end{array}\right) &=U_{\mathbf{k}}^{T}(t)\left(\begin{array}{c} d_{\mathbf{k}\sigma}\\ d_{\mathbf{k}+r\mathbf{Q},\sigma} \end{array}\right).
	\end{aligned}
\end{equation}

If $\mathbf{G}_{\mathbf{k}}$ denotes the Green's function in the quasiparticle basis and $\mathbf{\tilde{G}}_{\mathbf{k}}$ the corresponding Green's function in the physical electron basis, then
\begin{equation}
	\mathbf{\tilde{G}}_{\mathbf{k}}^{\gtrless}(t_{1},t_{2}) =U_{\mathbf{k}}^{T}(t_{1})\mathbf{G}_{\mathbf{k}}^{\gtrless}(t_{1},t_{2})U_{\mathbf{k}}^{*}(t_{2}),
\end{equation}
and the same transformation applies to the retarded, advanced, and Keldysh components because each is a linear combination of $G^{>}$ and $G^{<}$.

Writing
\begin{equation}
	U_{\mathbf{k}}(t)=U_{\mathbf{k}}^{0}+U_{\mathbf{k}}^{+}e^{-ir\Omega t}+U_{\mathbf{k}}^{-}e^{ir\Omega t},
\end{equation}
with
\begin{widetext}
	\begin{equation}
		U_{\mathbf{k}}^{0}=\left(\begin{array}{cc} u_{\mathbf{k}} & 0\\ 0 & u_{\mathbf{k}+r\mathbf{Q}} \end{array}\right),\quad U_{\mathbf{k}}^{+}=\left(\begin{array}{cc} 0 & v_{\mathbf{k}}\\ 0 & 0 \end{array}\right),\quad U_{\mathbf{k}}^{-}=\left(\begin{array}{cc} 0 & 0\\ v_{\mathbf{k}+r\mathbf{Q}} & 0 \end{array}\right),
	\end{equation}
	and expanding $\mathbf{G}_{\mathbf{k}}(t_{1},t_{2})$ in Floquet harmonics
	\begin{equation}
		\mathbf{G}_{\mathbf{k}}(t_{1},t_{2}) =\sum_{n}\int\frac{d\omega}{2\pi}e^{-i\omega t}e^{in\Omega\tau}\mathbf{G}_{\mathbf{k}}(\omega,n),
	\end{equation}
	one obtains the physical-basis Floquet components
	\begin{equation}
		\begin{split}
			\mathbf{\tilde{G}}_{\mathbf{k}}(\omega,n) &= U_{\mathbf{k}}^{0T}\mathbf{G}_{\mathbf{k}}(\omega,n)U_{\mathbf{k}}^{0*} +U_{\mathbf{k}}^{+T}\mathbf{G}_{\mathbf{k}}(\omega-r\Omega,n)U_{\mathbf{k}}^{+*} +U_{\mathbf{k}}^{-T}\mathbf{G}_{\mathbf{k}}(\omega+r\Omega,n)U_{\mathbf{k}}^{-*}\\
			&\quad +U_{\mathbf{k}}^{0T}\mathbf{G}_{\mathbf{k}}(\omega-r\Omega/2,n-r)U_{\mathbf{k}}^{+*} +U_{\mathbf{k}}^{+T}\mathbf{G}_{\mathbf{k}}(\omega-r\Omega/2,n+r)U_{\mathbf{k}}^{0*}\\
			&\quad +U_{\mathbf{k}}^{0T}\mathbf{G}_{\mathbf{k}}(\omega+r\Omega/2,n+r)U_{\mathbf{k}}^{-*} +U_{\mathbf{k}}^{-T}\mathbf{G}_{\mathbf{k}}(\omega+r\Omega/2,n-r)U_{\mathbf{k}}^{0*}\\
			&\quad +U_{\mathbf{k}}^{+T}\mathbf{G}_{\mathbf{k}}(\omega,n+2r)U_{\mathbf{k}}^{-*} +U_{\mathbf{k}}^{-T}\mathbf{G}_{\mathbf{k}}(\omega,n-2r)U_{\mathbf{k}}^{+*}.
		\end{split}
	\end{equation}
\end{widetext}
This is the quantity that enters the tunneling and transport expressions in the main text.
	
\section{Schwinger-Keldysh formulation and Dyson equation}\label{sec:sk-dyson}

This appendix summarizes the Schwinger-Keldysh structure used throughout the main text~\cite{Schwinger1961,Keldysh,Rammer86,Kamenev09,Stefanucci2013,HaugJauho2008}. Only the retarded, advanced, and Keldysh components are needed. In the Larkin-Ovchinnikov representation,
\begin{equation}
	\mathbf{G}=\left[\begin{array}{cc}G^{R}&G^{K}\\0&G^{A}\end{array}\right],\qquad \mathbf{\Sigma}=\left[\begin{array}{cc}\Sigma^{R}&\Sigma^{K}\\0&\Sigma^{A}\end{array}\right],
\end{equation}
with $G^{<}=(G^{K}-G^{R}+G^{A})/2$. Dyson's equation $\mathbf{G}=\mathbf{G}_{0}+\mathbf{G}_{0}\mathbf{\Sigma}\mathbf{G}$ implies
\begin{subequations}
	\begin{equation}
		G^{R/A}=G_{0}^{R/A}+G_{0}^{R/A}\Sigma^{R/A}G^{R/A},
	\end{equation}
	\begin{equation}
		G^{K}=\left(1+G^{R}\Sigma^{R}\right)G_{0}^{K}\left(1+\Sigma^{A}G^{A}\right)+G^{R}\Sigma^{K}G^{A}.
	\end{equation}
\end{subequations}

For the transport problem of Sec.~\ref{sec:transport}, Dyson's equation must be applied in the physical electron basis created by $\{f_{p}^{\dagger}\}$, because the tunneling Hamiltonian to the contacts is stationary in this basis. Denote by $\mathbf{\tilde{g}}$ the Green's functions of the isolated sliding CDW and by $\mathbf{\tilde{G}}$ the Green's functions after the contacts are attached. Then $\mathbf{\tilde{G}}=\mathbf{\tilde{g}}+\mathbf{\tilde{g}}\mathbf{\Sigma}\mathbf{\tilde{G}}$ becomes
\begin{subequations}
	\begin{equation}
		\mathbf{\tilde{G}}^{R/A}=\mathbf{\tilde{g}}^{R/A}+\mathbf{\tilde{g}}^{R/A}\Sigma^{R/A}\mathbf{\tilde{G}}^{R/A},
	\end{equation}
	\begin{equation}
		\mathbf{\tilde{G}}^{K}=\left(1+\mathbf{\tilde{G}}^{R}\Sigma^{R}\right)\mathbf{\tilde{g}}^{K}\left(1+\Sigma^{A}\mathbf{\tilde{G}}^{A}\right)+\mathbf{\tilde{G}}^{R}\Sigma^{K}\mathbf{\tilde{G}}^{A}.
	\end{equation}
\end{subequations}
When the periodic steady state is fixed by the contacts or by additional relaxation processes, the second term in the Keldysh equation is the relevant one, recovering the familiar relation $\mathbf{\tilde{G}}^{K}\simeq\mathbf{\tilde{G}}^{R}\Sigma^{K}\mathbf{\tilde{G}}^{A}$.

\section{Lead self-energy in the physical electron basis}\label{sec:lead-self-energy}

This appendix derives the stationary lead self-energy used in Sec.~\ref{sec:spectroscopy} and Sec.~\ref{sec:transport}. Let $p$ and $q$ label physical single-particle states in the CDW region. The contact Hamiltonian is
\begin{subequations}
	\begin{equation}
		H_{\alpha}=\sum_{k}\epsilon_{k\alpha}c_{k\alpha}^{\dagger}c_{k\alpha},
	\end{equation}
	\begin{equation}
		H_{T,\alpha}=\sum_{k,p,\sigma}V_{k\alpha,p}c_{k\alpha}^{\dagger}f_{p\sigma}+\mathrm{H.c.}
	\end{equation}
\end{subequations}
Because the couplings $V_{k\alpha,p}$ are time independent in the $f$ basis, the corresponding self-energy is stationary, with matrix elements
\begin{equation}
	\Sigma_{\alpha,pq}^{R(A,K)}(t_1,t_2)=\sum_{k}V_{k\alpha,p}^{*}g_{k\alpha}^{R(A,K)}(t_1-t_2)V_{k\alpha,q},
\end{equation}
or, after Fourier transformation,
\begin{equation}
	\Sigma_{\alpha,pq}^{R(A,K)}(\omega)=\sum_{k}V_{k\alpha,p}^{*}g_{k\alpha}^{R(A,K)}(\omega)V_{k\alpha,q}.
\end{equation}
For an equilibrium contact,
\begin{equation}
	\begin{split}
		g_{k\alpha}^{R/A}(\omega)&=\frac{1}{\omega-\epsilon_{k\alpha}\pm i0^{+}},\\
		g_{k\alpha}^{K}(\omega)&=2\pi i\left[2f_{\alpha}(\epsilon_{k\alpha})-1\right]\delta(\omega-\epsilon_{k\alpha}),
	\end{split}
\end{equation}
where $f_{\alpha}(\omega)=n_{F}(\omega-\mu_{\alpha})$. Therefore
\begin{subequations}
	\begin{equation}
		\Sigma_{\alpha,pq}^{R/A}(\omega)=\Lambda_{\alpha,pq}(\omega)\mp \frac{i}{2}\Gamma_{\alpha,pq}(\omega),
	\end{equation}
	\begin{equation}
		\Sigma_{\alpha,pq}^{K}(\omega)=i\left[2f_{\alpha}(\omega)-1\right]\Gamma_{\alpha,pq}(\omega),
	\end{equation}
\end{subequations}
with the linewidth matrix
\begin{equation}
	\Gamma_{\alpha,pq}(\omega)=2\pi\sum_{k}V_{k\alpha,p}^{*}V_{k\alpha,q}\delta(\omega-\epsilon_{k\alpha}).
\end{equation}
The total self-energy is $\Sigma=\sum_{\alpha}\Sigma_{\alpha}$. In the wide-band limit one sets $\Lambda_{\alpha}(\omega)\to 0$ and $\Gamma_{\alpha}(\omega)\to \Gamma_{\alpha}$.

\section{Time-averaging the current formula}\label{sec:Javg}

This appendix details the reduction of the double-time current expression to the stationary-contact form used in Eqs.~\eqref{eq:JPavg} and~\eqref{eq:JLavg} of the main text. The time-dependent current through contact $\alpha$ is
\begin{widetext}
	\begin{equation*}
		J_{\alpha}(t) = -\frac{2e}{\hbar}\int_{-\infty}^{t}dt_1 \mathrm{Im}\int\frac{d\epsilon}{2\pi}\mathrm{tr}\left\{e^{-i\epsilon(t_1-t)}\mathbf{\Gamma}^{\alpha}(\epsilon,t_{1},t) \left[\mathbf{\tilde{G}}^{<}(t,t_1)+f_{\alpha}(\epsilon)\mathbf{\tilde{G}}^{R}(t,t_1) \right] \right\},
	\end{equation*}
\end{widetext}
and the time-averaged current is
\begin{equation}
	\langle J_{\alpha}(t)\rangle = \lim_{T\to \infty}\frac{1}{T}\int_{-T/2}^{T/2}dt \,J_{\alpha}(t).
\end{equation}
The retarded term extends naturally over the full square domain because $\mathbf{\tilde{G}}^{R}(t,t_1)=0$ for $t_1>t$. For the lesser term, the symmetry $\mathbf{\tilde{G}}^{<(>)}(t_1,t_2)^{\dagger}=-\mathbf{\tilde{G}}^{<(>)}(t_2,t_1)$ together with $\mathbf{\Gamma}^{\alpha}(\epsilon,t_1,t_2)^{\dagger}=\mathbf{\Gamma}^{\alpha}(\epsilon,t_2,t_1)$ implies that the imaginary part of the integrand is symmetric under $t\leftrightarrow t_1$. The triangular domain can therefore be symmetrized, yielding
\begin{widetext}
	\begin{equation}
		\langle J_{\alpha}(t)\rangle = -\frac{e}{\hbar}\lim_{T\to\infty}\frac{1}{T}\int_{-T/2}^{T/2}dt\int_{-T/2}^{T/2}dt_1 \mathrm{Im}\int\frac{d\epsilon}{2\pi} \mathrm{tr}\left\{e^{-i\epsilon(t_1-t)}\mathbf{\Gamma}^{\alpha}(\epsilon,t_{1},t) \left[\mathbf{\tilde{G}}^{<}(t,t_1)+2f_{\alpha}(\epsilon)\mathbf{\tilde{G}}^{R}(t,t_1) \right] \right\}.
	\end{equation}
\end{widetext}
Introducing the relative time $s=t_{1}-t$ and the center-of-mass time $\tau=(t+t_{1})/2$, and using the double Fourier expansions
\begin{subequations}
	\begin{equation}
		\mathbf{\Gamma}^{\alpha}(\epsilon,t_{1},t)=\sum_{n}\int\frac{d\omega}{2\pi}e^{-i\omega s}e^{in\Omega\tau}\mathbf{\Gamma}^{\alpha}(\epsilon,\omega,n),
	\end{equation}
	\begin{equation}
		\mathbf{\tilde{G}}^{<(R)}(t,t_{1})=\sum_{l}\int\frac{d\omega^{\prime}}{2\pi}e^{i\omega^{\prime}s}e^{il\Omega\tau}\mathbf{\tilde{G}}^{<(R)}(\omega^{\prime},l),
	\end{equation}
\end{subequations}
substitution gives
\begin{widetext}
	\begin{equation}
		\begin{split}
			\langle J_{\alpha}(t)\rangle &= -\frac{e}{\hbar}\lim_{T\to\infty}\frac{1}{T}\sum_{n,l}\mathrm{Im}\int\frac{d\epsilon}{2\pi}\int\frac{d\omega}{2\pi}\int\frac{d\omega^{\prime}}{2\pi}\int d\tau\int ds\\
			&\quad \times e^{-i(\epsilon+\omega-\omega^{\prime})s}e^{i(n+l)\Omega\tau} \mathrm{tr}\left\{\mathbf{\Gamma}^{\alpha}(\epsilon,\omega,n)\left[\mathbf{\tilde{G}}^{<}(\omega^{\prime},l)+2f_{\alpha}(\epsilon)\mathbf{\tilde{G}}^{R}(\omega^{\prime},l)\right]\right\}.
		\end{split}
	\end{equation}
	Using
	\begin{equation}
		\lim_{T\to\infty}\frac{1}{T}\int_{-T/2}^{T/2}d\tau\,e^{i(n+l)\Omega\tau}=\delta_{n,-l},\qquad \int_{-\infty}^{\infty}ds\,e^{-i(\epsilon+\omega-\omega^{\prime})s}=2\pi\delta(\omega^{\prime}-\omega-\epsilon),
	\end{equation}
	the $\tau$, $s$, and $\omega^{\prime}$ integrations yield
	\begin{equation}
	\langle J_{\alpha}(t)\rangle = -\frac{e}{\hbar}\sum_{n}\mathrm{Im}\int\frac{d\omega}{2\pi}\int\frac{d\epsilon}{2\pi}\mathrm{tr}\left\{\mathbf{\Gamma}^{\alpha}(\epsilon,\omega,n) \left[\mathbf{\tilde{G}}^{<}(\omega+\epsilon,-n)+2f_{\alpha}(\epsilon)\mathbf{\tilde{G}}^{R}(\omega+\epsilon,-n) \right] \right\}.
	\end{equation}
\end{widetext}
For a stationary contact, $\mathbf{\Gamma}^{\alpha}(\epsilon,t_1,t_2)$ is independent of the center time and only the $n=0$ harmonic survives. The expression then reduces to
\begin{equation}\label{eq:JavgStationary}
\langle J_{\alpha}\rangle = -\frac{e}{\hbar}\mathrm{Im}\int\frac{d\epsilon}{2\pi}\mathrm{tr}\left\{\mathbf{\Gamma}^{\alpha}(\epsilon)\left[\mathbf{\tilde{G}}^{<}(\epsilon,0)+2f_{\alpha}(\epsilon)\mathbf{\tilde{G}}^{R}(\epsilon,0)\right]\right\},
\end{equation}
which is the form used in Eqs.~\eqref{eq:JPavg} and~\eqref{eq:JLavg} of the main text.
	
\section{Single-contact tunneling spectra and inverse-current oscillations \texorpdfstring{$dI/dV$}{dI/dV}}\label{sec:realistic-single-contact}

This appendix provides the explicit formulas underlying Figs.~\ref{Fig:inverseI} and~\ref{Fig:tunnelSpec} of the main text, and demonstrates that the near-edge spectra at fixed current and the fixed-bias inverse-current oscillation are two complementary cuts of the same Floquet sideband ladder.

For the weak-probe spectroscopy of Sec.~\ref{sec:spectroscopy}, the normalized differential conductance is approximated by
\begin{equation}
	\frac{G(V,I)}{G_{N}} \simeq \int dE\,A_{\mathrm{loc}}^{\mathrm{ave}}(E;I)\left[-\frac{\partial n_{F}(E-eV)}{\partial E}\right].
\end{equation}
To regularize the square-root singularity at the CDW gap edge, we introduce a Dynes broadening $\gamma$ and write
\begin{equation}
	D_{\gamma}(E)=\mathrm{Re}\frac{E+i\gamma}{\sqrt{(E+i\gamma)^{2}-\Delta^{2}}},
\end{equation}
so that the time-averaged spectrum in the sliding state is
\begin{equation}
	A_{\mathrm{loc}}^{\mathrm{ave}}(E;I)\simeq \frac{1}{2}\left[D_{\gamma}\left(E+\frac{\hbar\Omega(I)}{2}\right)+D_{\gamma}\left(E-\frac{\hbar\Omega(I)}{2}\right)\right].
\end{equation}
This is the near-edge envelope used in Fig.~\ref{Fig:tunnelSpec}: it captures the split singularities of the sliding-CDW DOS while retaining a phenomenological broadening suitable for comparison with experiment.

For the (TaSe$_4$)$_2$I nanowire data of Ref.~\cite{Le24}, we take $\Delta=165\,\mathrm{meV}$, temperatures $T=1.55\,\mathrm{K}$ and $8\,\mathrm{K}$, and the experimental oscillation period $\Delta(1/I)=0.00269\,\mu\mathrm{A}^{-1}$. These values fix the product
\begin{equation}
	N_{\mathrm{eff}}\,\epsilon=\frac{h}{2e\,\Delta(1/I)}\simeq 4.80\,\mathrm{eV},
\end{equation}
where $\epsilon=\mu-E_{v}$ is the contact mismatch between the probe chemical potential and the lower CDW gap edge, and $N_{\mathrm{eff}}$ denotes the effective number of tunneling channels participating coherently in the spectroscopy. Therefore
\begin{equation}
	N_{\mathrm{eff}}\simeq \frac{4.80\,\mathrm{eV}}{\epsilon},\qquad \hbar\Omega[\mathrm{meV}] \simeq 0.00269\,I[\mu\mathrm{A}]\,\epsilon[\mathrm{meV}].
\end{equation}
For the reference choice $\epsilon=10\,\mathrm{meV}$ used in Fig.~\ref{Fig:tunnelSpec}, this gives $N_{\mathrm{eff}}\simeq 4.8\times 10^{2}$, together with $\hbar\Omega=0.269$, $0.538$, and $1.345\,\mathrm{meV}$ at $I=10$, $20$, and $50\,\mu\mathrm{A}$, respectively. For $\epsilon=5$, $10$, and $20\,\mathrm{meV}$, the corresponding channel numbers are about $9.6\times 10^{2}$, $4.8\times 10^{2}$, and $2.4\times 10^{2}$. With $\gamma=0.30\,\mathrm{meV}$ and thermal widths $3.5k_{B}T=0.47\,\mathrm{meV}$ at $1.55\,\mathrm{K}$ and $2.41\,\mathrm{meV}$ at $8\,\mathrm{K}$, the reference value $\hbar\Omega(20\,\mu\mathrm{A})\simeq0.54\,\mathrm{meV}$ is barely resolved at base temperature and strongly smeared at $8\,\mathrm{K}$, consistent with Fig.~\ref{Fig:tunnelSpec}.

To connect these fixed-current spectra with the observed $1/I$ oscillation, the full Floquet sideband ladder must be restored. Because $A_{n}(E)=A_{0}(E-n\hbar\Omega)$ and $A_{0}(E)$ has edge singularities at $E_{v}\pm\hbar\Omega/2$, the positive-bias singularities occur at
\begin{equation}
	E_{v}+\left(m+\frac{1}{2}\right)\hbar\Omega,\qquad m=0,1,2,\ldots.
\end{equation}
At a fixed bias offset $\epsilon=eV-E_{v}$, the oscillatory part of the single-contact response is governed by repeated crossings of these edge singularities. A convenient way to isolate the singular contribution of one broadened positive edge is
\begin{equation}\label{eq:deltaGedge}
	\delta G_{\mathrm{edge}}(\delta;T,\gamma)=\int dE\,\left[D_{\gamma}(E)-1\right]\left[-\frac{\partial n_{F}(E-\Delta-\delta)}{\partial E}\right],
\end{equation}
where $\delta$ measures detuning from the positive gap edge and subtraction of $1$ removes the smooth normal-state background. Repeating this near-edge profile along the sideband ladder gives the fixed-bias oscillatory contribution
\begin{equation}\label{eq:fixedBiasOscillation}
	\delta G_{\mathrm{osc}}(I,\epsilon;T)\simeq \sum_{m=0}^{\infty}\delta G_{\mathrm{edge}}\!\left(\epsilon-\left(m+\frac{1}{2}\right)\hbar\Omega(I);T,\gamma\right).
\end{equation}
Equation~\eqref{eq:fixedBiasOscillation} omits the smooth background and possible sideband-dependent matrix elements, but provides a direct calculation of the singular oscillatory part generated by the Floquet ladder. The resonance condition reads
\begin{equation}\label{eq:fixedBiasCondition}
	\epsilon=\left(m+\frac{1}{2}\right)\hbar\Omega(I_{m})=\left(m+\frac{1}{2}\right)\frac{hI_{m}}{2eN_{\mathrm{eff}}},
\end{equation}
which gives
\begin{equation}\label{eq:fixedBiasPeriod}
	\frac{1}{I_{m}}=\frac{h}{2eN_{\mathrm{eff}}\epsilon}\left(m+\frac{1}{2}\right),\qquad \Delta(1/I)=\frac{h}{2eN_{\mathrm{eff}}\epsilon}.
\end{equation}
For $\epsilon=10\,\mathrm{meV}$, Eq.~\eqref{eq:fixedBiasPeriod} yields $\Delta(1/I)=0.00269\,\mu\mathrm{A}^{-1}$, exactly the experimental value of Ref.~\cite{Le24}. The main-text Fig.~\ref{Fig:inverseI} shows the corresponding fixed-bias oscillatory contribution evaluated from Eq.~\eqref{eq:fixedBiasOscillation} with the same reference broadening $\gamma=0.30\,\mathrm{meV}$ used in Fig.~\ref{Fig:tunnelSpec}. The peaks are equally spaced in $1/I$, and the higher temperature suppresses and smears the oscillatory sideband structure through the thermal convolution in Eq.~\eqref{eq:deltaGedge}.

It is instructive to compare $N_{\mathrm{eff}}$ with a crude geometric estimate. Reference~\cite{Le24} reports a typical cross section smaller than $0.03\,\mu\mathrm{m}^{2}$ for device D1. Taking one effective chain per $\mathrm{nm}^{2}$ gives a geometric channel count of order $N_{\mathrm{geom}}\sim 3\times 10^{4}$, two orders of magnitude larger than the $N_{\mathrm{eff}}\sim 4.8\times 10^{2}$ extracted above. The measured period would then imply $\epsilon\sim 0.16\,\mathrm{meV}$ and $\hbar\Omega(20\,\mu\mathrm{A})\sim 8.6\times 10^{-3}\,\mathrm{meV}$, far below both $\gamma$ and $3.5k_{B}T$. The experimentally relevant spectroscopy signal therefore cannot be governed by the full geometric chain count; instead it is naturally controlled by a much smaller contact-local effective channel number, providing quantitative support for the percolating filament picture developed in Sec.~\ref{sec:spectroscopy}.

\section{Current-driven segmented transport for terminal-pair voltages}\label{sec:segmented-transport}

This appendix implements the measurement logic of Ref.~\cite{Le24} at the minimal effective level used in Fig.~\ref{Fig:segmentedTransport}. There a persistent current source fixes one common through-current in the whole device, while the voltage is measured between selected terminal pairs. To model that geometry, we keep the oscillatory kernel generated by the Floquet sideband ladder of Appendix~\ref{sec:realistic-single-contact}, but distribute the measured voltage over three series segments $1$-$2$, $2$-$3$, and $3$-$4$. The label ``current driven'' is used deliberately: the theory takes the net through-current as the control parameter and assigns the terminal voltages through an effective series decomposition, which is the steady-state idealization of the experimental persistent-current setup.

Define the normalized coherent oscillatory kernel
\begin{equation}\label{eq:segmentedKernel}
	\mathcal{O}_{\mathrm{osc}}(I;T)=\frac{1}{\mathcal{N}_{T}}\sum_{m=0}^{\infty}\delta G_{\mathrm{edge}}\!\left(\epsilon-\left(m+\frac{1}{2}\right)\hbar\Omega(I);T,\gamma\right),
\end{equation}
where $\delta G_{\mathrm{edge}}$ is the broadened edge profile of Eq.~\eqref{eq:deltaGedge}, $\epsilon=10\,\mathrm{meV}$ is the reference contact mismatch, and the normalization $\mathcal{N}_{T}$ ensures $\max_{I}|\mathcal{O}_{\mathrm{osc}}(I;T)|=1$ within the plotted current window $10\leq I\leq 50\,\mu\mathrm{A}$. Equation~\eqref{eq:segmentedKernel} inherits the same inverse-current spacing $\Delta(1/I)=0.00269\,\mu\mathrm{A}^{-1}$ as Eq.~\eqref{eq:fixedBiasPeriod}; only its assignment to different voltage probes is changed here.

The three segment voltages are written as
\begin{equation}\label{eq:segmentedVoltages}
	\begin{aligned}
		V_{12}(I;T) &= V_{12}^{\mathrm{bg}}(I)-A_{12}(T)\mathcal{O}_{\mathrm{osc}}(I;T),\\
		V_{23}(I;T) &= V_{23}^{\mathrm{bg}}(I)-A_{23}(T)\mathcal{O}_{\mathrm{osc}}(I;T),\\
		V_{34}(I;T) &= V_{34}^{\mathrm{bg}}(I)-A_{34}(T)\mathcal{O}_{\mathrm{osc}}(I;T),
	\end{aligned}
\end{equation}
with smooth background voltages $V_{ij}^{\mathrm{bg}}(I)$ and temperature-dependent oscillation amplitudes $A_{ij}(T)$. For current driven between terminals $1$ and $4$, the measured inner- and outer-terminal voltages are
\begin{equation}\label{eq:segmentedMeasured}
	\begin{aligned}
		V_{2-3}^{1-4}(I)&=V_{23}(I;T),\\
		V_{1-4}^{1-4}(I)&=V_{12}(I;T)+V_{23}(I;T)+V_{34}(I;T).
	\end{aligned}
\end{equation}

Figure~\ref{Fig:segmentedTransport} uses the minimal visibility choice
\begin{equation}
\begin{split}
& A_{12}(T)=A_{34}(T)=0, \\
& A_{23}(1.55\,\mathrm{K})=0.90\,\mathrm{mV},\\ 
& A_{23}(8\,\mathrm{K})=0.22\,\mathrm{mV},
\end{split}
\end{equation}
corresponding to a coherent inner segment together with effectively dephased outer segments. The smooth backgrounds are taken as representative low-order polynomials,
\begin{equation}
	V_{12}^{\mathrm{bg}}(I)=16.0+0.22\,I+0.0025(I-24)^2\ \mathrm{mV},
\end{equation}
\begin{equation}
	V_{23}^{\mathrm{bg}}(I)=2.5+0.030\,I+0.0008(I-24)^2\ \mathrm{mV},
\end{equation}
\begin{equation}
	V_{34}^{\mathrm{bg}}(I)=V_{12}^{\mathrm{bg}}(I),
\end{equation}
with $I$ measured in $\mu\mathrm{A}$. These background terms are not universal microscopic predictions; they simply represent the smooth series voltage of the outer regions. In this minimal limit the raw outer-terminal voltage is dominated by $V_{12}^{\mathrm{bg}}+V_{34}^{\mathrm{bg}}$, while the coherent $1/I$ oscillation is inherited from the inner segment through $V_{23}$. Subtracting the smooth background from $V_{1-4}^{1-4}(I)$ therefore recovers the same oscillatory kernel that is directly visible in $V_{2-3}^{1-4}(I)$. The segmented model thus cleanly separates two ingredients that are entangled in the experiment: one common current fixes the Floquet frequency, while spatially nonuniform coherence determines the terminal dependence of the observed visibility.

\section{Near-threshold two-terminal conductance at fixed sliding frequency}\label{sec:near-threshold-two-terminal}

This appendix provides the homogeneous reference limit used to generate Fig.~\ref{Fig:twoTerminal}. The goal is not to reproduce the persistent-current-driven experiment of Ref.~\cite{Le24}, but to show how the Floquet sideband ladder appears when a finite sliding span is voltage biased between two reservoirs. For the genuine two-terminal device, the stationary-contact current of Eq.~\eqref{eq:JavgStationary} can be recast into the usual Floquet-Landauer form for a noninteracting periodically driven region,
\begin{equation}\label{eq:twoTerminalLandauer}
	\overline{I}(V,\Omega)=\frac{e}{h}\sum_{n}\int dE\,T_{n}(E;\Omega)\left[f_{L}(E)-f_{R}(E)\right],
\end{equation}
where $T_{n}(E;\Omega)$ is the transmission probability from an incoming state at energy $E$ to an outgoing state at $E+n\hbar\Omega$. A full microscopic evaluation of $T_{n}$ for a finite depinned device is beyond the present parameter-estimate calculation, because one must specify the finite sliding region, its channel count, and the self-consistent relation between $\Omega$ and the through-current. Here we keep $\hbar\Omega$ as an externally specified sliding frequency and isolate the near-threshold structure controlled by the Floquet gap edges.

Near the positive gap edge, the same broadened singular profile used in the single-contact analysis provides the line shape of one threshold,
\begin{equation}\label{eq:twoTerminalEdge}
	g_{\mathrm{edge}}(\delta;T,\gamma)=\int dE\,D_{\gamma}(E)\left[-\frac{\partial n_{F}(E-\Delta-\delta)}{\partial E}\right],
\end{equation}
where $\delta$ measures detuning from the positive edge. In a symmetric two-terminal bias window, $dI/dV$ near $eV\simeq 2\Delta$ is controlled by the transmission at the edges of the bias window, $E\simeq\pm eV/2$. The Floquet ladder therefore produces inward threshold replicas at
\begin{equation}\label{eq:twoTerminalThreshold}
	E_{m}=\Delta-\left(m+\frac{1}{2}\right)\hbar\Omega,\qquad eV_{m}=2E_{m}=2\Delta-(2m+1)\hbar\Omega.
\end{equation}
Keeping only the thresholds inside a narrow near-threshold voltage window, the differential conductance is estimated as
\begin{equation}\label{eq:twoTerminalNearThreshold}
	\frac{dI}{dV}(V,\Omega;T)\propto\sum_{m\in\mathcal{M}_{\mathrm{vis}}} g_{\mathrm{edge}}\!\left(\frac{eV}{2}-\Delta+\left(m+\frac{1}{2}\right)\hbar\Omega;T,\gamma\right),
\end{equation}
where $\mathcal{M}_{\mathrm{vis}}$ denotes the few inward thresholds visible in the plotted voltage range. Equation~\eqref{eq:twoTerminalNearThreshold} is the two-terminal counterpart of the single-contact sideband sum: instead of scanning current at fixed bias, one scans bias at fixed $\hbar\Omega$, so the ladder appears as threshold shifts linear in $\hbar\Omega$.

Figure~\ref{Fig:twoTerminal} evaluates Eq.~\eqref{eq:twoTerminalNearThreshold} with the realistic reference parameters $\Delta=165\,\mathrm{meV}$, $\gamma=0.30\,\mathrm{meV}$, and $T=1.55\,\mathrm{K}$ or $8\,\mathrm{K}$. The representative values $\hbar\Omega=0.269$, $0.538$, and $1.345\,\mathrm{meV}$ are the same sliding frequencies used in the single-contact estimate, but here treated as fixed inputs rather than tied self-consistently to a through-current. The low-temperature curves display weak shoulders below $2\Delta$ at the thresholds of Eq.~\eqref{eq:twoTerminalThreshold}, while the high-temperature curves show that these structures are readily washed out once $3.5k_{B}T$ exceeds the sideband spacing. The homogeneous two-terminal case is thus naturally discussed in terms of threshold lines in the $(V,\Omega)$ plane; converting those lines into the terminal-pair voltages of the experiment requires the current-driven construction of Appendix~\ref{sec:segmented-transport}.

	\bibliography{OCDW}

@article{Floquet,
  author = {Floquet, G.},
  title = {Sur les \'equations diff\'erentielles lin\'eaires \`a coefficients p\'eriodiques},
  journal = {Annales scientifiques de l'\'Ecole Normale Sup\'erieure},
  series = {2e s{\'e}rie},
  volume = {12},
  pages = {47--88},
  year = {1883},
  doi = {10.24033/asens.220},
  url = {http://www.numdam.org/articles/10.24033/asens.220/}
}

@article{Schwinger1961,
  title = {Brownian Motion of a Quantum Oscillator},
  author = {Schwinger, Julian S.},
  journal = {Journal of Mathematical Physics},
  year = {1961},
  volume = {2},
  pages = {407--432},
  doi = {10.1063/1.1703727}
}

@article{Keldysh,
  author = {Keldysh, L. V.},
  journal = {Sov. Phys. JETP},
  pages = {1018--1026},
  title = {Diagram technique for nonequilibrium processes},
  volume = {20},
  year = {1965}
}

@article{Kamenev09,
  author = {Kamenev, Alex and Levchenko, Alex},
  title = {Keldysh technique and non-linear $\sigma$-model: basic principles and applications},
  journal = {Advances in Physics},
  volume = {58},
  number = {3},
  pages = {197--319},
  year = {2009},
  publisher = {Taylor \& Francis},
  doi = {10.1080/00018730902850504},
  url = {https://doi.org/10.1080/00018730902850504}
}

@article{Grifoni,
  title = {Driven quantum tunneling},
  journal = {Physics Reports},
  volume = {304},
  number = {5},
  pages = {229--354},
  year = {1998},
  issn = {0370-1573},
  doi = {10.1016/S0370-1573(98)00022-2},
  url = {https://www.sciencedirect.com/science/article/pii/S0370157398000222},
  author = {Milena Grifoni and Peter H\"anggi}
}

@article{Platero04,
  title = {Photon-assisted transport in semiconductor nanostructures},
  journal = {Physics Reports},
  volume = {395},
  number = {1},
  pages = {1--157},
  year = {2004},
  issn = {0370-1573},
  doi = {10.1016/j.physrep.2004.01.004},
  url = {https://www.sciencedirect.com/science/article/pii/S0370157304000304},
  author = {Gloria Platero and Ram\'on Aguado}
}

@article{Jauho94,
  title = {Time-dependent transport in interacting and noninteracting resonant-tunneling systems},
  author = {Jauho, Antti-Pekka and Wingreen, Ned S. and Meir, Yigal},
  journal = {Phys. Rev. B},
  volume = {50},
  issue = {8},
  pages = {5528--5544},
  year = {1994},
  month = {Aug},
  publisher = {American Physical Society},
  doi = {10.1103/PhysRevB.50.5528},
  url = {https://link.aps.org/doi/10.1103/PhysRevB.50.5528}
}

@article{Kohler05,
  title = {Driven quantum transport on the nanoscale},
  journal = {Physics Reports},
  volume = {406},
  number = {6},
  pages = {379--443},
  year = {2005},
  issn = {0370-1573},
  doi = {10.1016/j.physrep.2004.11.002},
  url = {https://www.sciencedirect.com/science/article/pii/S0370157304005071},
  author = {Sigmund Kohler and J\"org Lehmann and Peter H\"anggi}
}

@article{SDW94,
  title = {The dynamics of spin-density waves},
  author = {Gr\"uner, G.},
  journal = {Rev. Mod. Phys.},
  volume = {66},
  issue = {1},
  pages = {1--24},
  year = {1994},
  month = {Jan},
  publisher = {American Physical Society},
  doi = {10.1103/RevModPhys.66.1},
  url = {https://link.aps.org/doi/10.1103/RevModPhys.66.1}
}

@article{SDW60,
  title = {Giant Spin Density Waves},
  author = {Overhauser, A. W.},
  journal = {Phys. Rev. Lett.},
  volume = {4},
  issue = {9},
  pages = {462--465},
  year = {1960},
  month = {May},
  publisher = {American Physical Society},
  doi = {10.1103/PhysRevLett.4.462},
  url = {https://link.aps.org/doi/10.1103/PhysRevLett.4.462}
}

@article{SDW62,
  title = {Spin Density Waves in an Electron Gas},
  author = {Overhauser, A. W.},
  journal = {Phys. Rev.},
  volume = {128},
  issue = {3},
  pages = {1437--1452},
  year = {1962},
  month = {Nov},
  publisher = {American Physical Society},
  doi = {10.1103/PhysRev.128.1437},
  url = {https://link.aps.org/doi/10.1103/PhysRev.128.1437}
}

@article{CDW88,
  title = {The dynamics of charge-density waves},
  author = {Gr\"uner, G.},
  journal = {Rev. Mod. Phys.},
  volume = {60},
  issue = {4},
  pages = {1129--1181},
  year = {1988},
  month = {Oct},
  publisher = {American Physical Society},
  doi = {10.1103/RevModPhys.60.1129},
  url = {https://link.aps.org/doi/10.1103/RevModPhys.60.1129}
}

@book{Peierls,
  title = {Quantum Theory of Solids},
  author = {Peierls, R. E.},
  publisher = {Oxford University Press},
  address = {Oxford},
  year = {1955}
}

@article{Frohlich54,
  author = {Fr\"ohlich, Herbert},
  title = {On the theory of superconductivity: the one-dimensional case},
  journal = {Proceedings of the Royal Society of London. Series A. Mathematical and Physical Sciences},
  volume = {223},
  number = {1154},
  pages = {296--305},
  year = {1954},
  doi = {10.1098/rspa.1954.0116}
}

@book{Mahan,
  address = {New York},
  author = {Mahan, G. D.},
  publisher = {Plenum},
  title = {Many-Particle Physics},
  year = {2000},
  edition = {3}
}

@article{Uhrig,
  title = {Positivity of the Spectral Densities of Retarded Floquet Green Functions},
  author = {Uhrig, G\"otz S. and Kalthoff, Mona H. and Freericks, James K.},
  journal = {Phys. Rev. Lett.},
  volume = {122},
  issue = {13},
  pages = {130604},
  year = {2019},
  month = {Apr},
  publisher = {American Physical Society},
  doi = {10.1103/PhysRevLett.122.130604},
  url = {https://link.aps.org/doi/10.1103/PhysRevLett.122.130604}
}

@article{Le24,
  author = {Le, Tian and Jiang, Ruiyang and Tu, Linfeng and Bian, Renji and Ma, Yiwen and Shi, Yunteng and Jia, Ke and Li, Zhilin and Lyu, Zhaozheng and Cao, Xuewei and Shen, Jie and Liu, Guangtong and Shi, Youguo and Liu, Fucai and Zhou, Yi and Lu, Li and Qu, Fanming},
  title = {Inverse-current quantum electro-oscillations in a charge-density wave insulator},
  journal = {Phys. Rev. B},
  volume = {109},
  issue = {24},
  pages = {245123},
  year = {2024},
  month = {Jun},
  publisher = {American Physical Society},
  doi = {10.1103/PhysRevB.109.245123},
  url = {https://link.aps.org/doi/10.1103/PhysRevB.109.245123}
}

@article{Sambe,
  title = {Steady States and Quasienergies of a Quantum-Mechanical System in an Oscillating Field},
  author = {Sambe, Hideo},
  journal = {Phys. Rev. A},
  volume = {7},
  issue = {6},
  pages = {2203--2213},
  year = {1973},
  month = {Jun},
  publisher = {American Physical Society},
  doi = {10.1103/PhysRevA.7.2203},
  url = {https://link.aps.org/doi/10.1103/PhysRevA.7.2203}
}

@article{Shirley,
  title = {Solution of the Schr\"odinger Equation with a Hamiltonian Periodic in Time},
  author = {Shirley, Jon H.},
  journal = {Phys. Rev.},
  volume = {138},
  issue = {4B},
  pages = {B979--B987},
  year = {1965},
  month = {May},
  publisher = {American Physical Society},
  doi = {10.1103/PhysRev.138.B979},
  url = {https://link.aps.org/doi/10.1103/PhysRev.138.B979}
}

@article{Rammer86,
  title = {Quantum field-theoretical methods in transport theory of metals},
  author = {Rammer, J. and Smith, H.},
  journal = {Rev. Mod. Phys.},
  volume = {58},
  issue = {2},
  pages = {323--359},
  year = {1986},
  month = {Apr},
  publisher = {American Physical Society},
  doi = {10.1103/RevModPhys.58.323},
  url = {https://link.aps.org/doi/10.1103/RevModPhys.58.323}
}

@article{OkaKitamura2019,
  author = {Oka, Takashi and Kitamura, Sota},
  title = {Floquet Engineering of Quantum Materials},
  journal = {Annu. Rev. Condens. Matter Phys.},
  volume = {10},
  pages = {387--408},
  year = {2019},
  doi = {10.1146/annurev-conmatphys-031218-013423}
}

@article{RudnerLindner2020,
  author = {Rudner, Mark S. and Lindner, Netanel H.},
  title = {Band structure engineering and non-equilibrium dynamics in Floquet topological insulators},
  journal = {Nat. Rev. Phys.},
  volume = {2},
  pages = {229--244},
  year = {2020},
  doi = {10.1038/s42254-020-0149-7}
}

@article{Bukov2015,
  author = {Bukov, Marin and D'Alessio, Luca and Polkovnikov, Anatoli},
  title = {Universal high-frequency behavior of periodically driven systems: from dynamical stabilization to Floquet engineering},
  journal = {Adv. Phys.},
  volume = {64},
  number = {2},
  pages = {139--226},
  year = {2015},
  publisher = {Taylor \& Francis},
  doi = {10.1080/00018732.2015.1055918}
}

@book{Stefanucci2013,
  author = {Stefanucci, Gianluca and van Leeuwen, Robert},
  title = {Nonequilibrium Many-Body Theory of Quantum Systems: A Modern Introduction},
  publisher = {Cambridge University Press},
  address = {Cambridge},
  year = {2013},
  doi = {10.1017/CBO9781139023979}
}

@book{HaugJauho2008,
  author = {Haug, Hartmut and Jauho, Antti-Pekka},
  title = {Quantum Kinetics in Transport and Optics of Semiconductors},
  edition = {2},
  series = {Springer Series in Solid-State Sciences},
  volume = {123},
  publisher = {Springer},
  address = {Berlin},
  year = {2008},
  doi = {10.1007/978-3-540-73564-2}
}

@article{Buttiker1986,
  author = {B\"uttiker, Markus},
  title = {Four-Terminal Phase-Coherent Conductance},
  journal = {Phys. Rev. Lett.},
  volume = {57},
  issue = {14},
  pages = {1761--1764},
  year = {1986},
  doi = {10.1103/PhysRevLett.57.1761}
}

@article{LeeRiceAnderson1974,
  author = {Lee, P. A. and Rice, T. M. and Anderson, P. W.},
  title = {Conductivity from charge or spin density waves},
  journal = {Solid State Commun.},
  volume = {14},
  pages = {703--709},
  year = {1974},
  doi = {10.1016/0038-1098(74)90341-3}
}

@article{FukuyamaLee1978,
  author = {Fukuyama, H. and Lee, P. A.},
  title = {Dynamics of the charge-density wave. {I}. {Impurity} pinning in a single chain},
  journal = {Phys. Rev. B},
  volume = {17},
  pages = {535--541},
  year = {1978},
  doi = {10.1103/PhysRevB.17.535}
}

@article{LeeRice1979,
  author = {Lee, P. A. and Rice, T. M.},
  title = {Electric field depinning of charge density waves},
  journal = {Phys. Rev. B},
  volume = {19},
  pages = {3970--3980},
  year = {1979},
  doi = {10.1103/PhysRevB.19.3970}
}

@article{Bardeen1979,
  author = {Bardeen, John},
  title = {Theory of non-{Ohmic} conduction from charge-density waves in {NbSe3}},
  journal = {Phys. Rev. Lett.},
  volume = {42},
  issue = {22},
  pages = {1498--1500},
  year = {1979},
  doi = {10.1103/PhysRevLett.42.1498}
}

@article{Monceau1976,
  author = {Monceau, P. and Ong, N. P. and Portis, A. M. and Meerschaut, A. and Rouxel, J.},
  title = {Electric-field breakdown of charge-density-wave-induced anomalies in {NbSe3}},
  journal = {Phys. Rev. Lett.},
  volume = {37},
  issue = {10},
  pages = {602--606},
  year = {1976},
  doi = {10.1103/PhysRevLett.37.602}
}

@article{FlemingGrimes1979,
  author = {Fleming, R. M. and Grimes, C. C.},
  title = {Sliding-mode conductivity in {NbSe3}: {Observation} of a threshold electric field and conduction noise},
  journal = {Phys. Rev. Lett.},
  volume = {42},
  issue = {21},
  pages = {1423--1426},
  year = {1979},
  doi = {10.1103/PhysRevLett.42.1423}
}

@article{Monceau2012,
  author = {Monceau, Pierre},
  title = {Electronic crystals: an experimental overview},
  journal = {Adv. Phys.},
  volume = {61},
  number = {4},
  pages = {325--424},
  year = {2012},
  publisher = {Taylor \& Francis},
  doi = {10.1080/00018732.2012.748709}
}

@book{Imry2002,
  author = {Imry, Yoseph},
  title = {Introduction to Mesoscopic Physics},
  edition = {2},
  publisher = {Oxford University Press},
  address = {Oxford},
  year = {2002},
  isbn = {978-0-19-850738-3},
  doi = {10.1093/oso/9780198507383.001.0001}
}

@book{LarkinVarlamov2009,
  author = {Larkin, A. I. and Varlamov, A. A.},
  title = {Theory of Fluctuations in Superconductors},
  edition = {Revised},
  series = {International Series of Monographs on Physics},
  volume = {127},
  publisher = {Oxford University Press},
  address = {Oxford},
  year = {2009},
  isbn = {978-0-19-956299-2}
}

@misc{yizhou76-sudo2026SlidingCDW,
  author = {yizhou76-sudo},
  title = {Sliding-CDW: Documents for Sliding CDW Theory},
  year = {2026},
  howpublished = {\url{https://github.com/yizhou76-sudo/Sliding-CDW}},
  note = {Accessed: 2026-05-07}
}
	
\end{document}